\documentclass[twocolumn,superscriptaddress,pra]{revtex4-1}
\usepackage{graphicx}
\usepackage[retainorgcmds]{IEEEtrantools}
\usepackage{amsmath}
\usepackage{latexsym}
\usepackage{amsfonts}
\usepackage{amssymb}
\usepackage{array}
\usepackage{epsfig}
\usepackage{color}
\usepackage{nicefrac}
\usepackage{hyperref}
\usepackage{braket}
\usepackage{times,txfonts}
\usepackage[table]{xcolor}
\usepackage{tabularx}
\usepackage{enumitem}

\DeclareMathOperator{\diag}{diag}
\newcommand{\ud}{\,\mathrm{d}}
\DeclareMathOperator{\tr}{tr}

\newcolumntype{b}{X}
\newcolumntype{s}{>{\hsize=.3\hsize}X}
\newcolumntype{m}{>{\hsize=.08\hsize}X}

\begin{document}

\title{Reconciliation of quantum local master equations with thermodynamics}

\author{Gabriele De Chiara}
\affiliation{Centre  for  Theoretical  Atomic,  Molecular  and  Optical  Physics, Queen's  University  Belfast,  Belfast  BT7 1NN,  United  Kingdom}
\affiliation{Kavli Institute of Theoretical Physics (KITP), University of California, Santa Barbara CA 93106-4030, United States of America}
\author{Gabriel Landi}
\affiliation{Instituto de F\'isica da Universidade de S\~ao Paulo,  05314-970 S\~ao Paulo, Brazil}

\author{Adam Hewgill}
\author{Brendan Reid}
\author{Alessandro Ferraro}
\affiliation{Centre  for  Theoretical  Atomic,  Molecular  and  Optical  Physics, Queen's  University  Belfast,  Belfast  BT7 1NN,  United  Kingdom}

\author{Augusto J. Roncaglia}
\affiliation{Departamento de F\'isica, Facultad de Ciencias Exactas y Naturales, Universidad de Buenos Aires and IFIBA, CONICET, Ciudad Universitaria, 1428 Buenos Aires, Argentina}
\author{Mauro Antezza}
\address{Laboratoire Charles Coulomb (L2C), UMR 5221 CNRS-Universit\'{e} de Montpellier, F-34095 Montpellier, France}
\address{Institut Universitaire de France, 1 rue Descartes, F-75231 Paris Cedex 05, France}
\affiliation{Kavli Institute of Theoretical Physics (KITP), University of California, Santa Barbara CA 93106-4030, United States of America}

\begin{abstract}
The study of open quantum systems often relies on approximate master equations derived under the assumptions of weak coupling to the environment. However when the system is made of several interacting subsystems such a derivation is in many cases very hard. An alternative method, employed especially in the modelling of transport in mesoscopic systems, consists in using {\it local} master equations containing Lindblad operators acting locally only on the corresponding subsystem. It has been shown that this approach however generates inconsistencies with the laws of thermodynamics. In this paper we demonstrate that using a microscopic model of local master equations based on repeated collisions all thermodynamic inconsistencies can be resolved by correctly taking into account the breaking of global detailed balance related to the work cost of maintaining the collisions. We provide examples based on a chain of quantum harmonic oscillators whose ends are connected to thermal reservoirs at different temperatures. We prove that this system behaves precisely as a quantum heat engine or refrigerator, with properties that are fully consistent with basic thermodynamics.
\end{abstract}

\maketitle



%
%
\section{\label{sec:int}Introduction}
%
%

The description and manipulation of energy transfer at the quantum scale is a problem of fundamental and technological importance, receiving increasing attention in recent years. On one side, the study of energy transport through mesoscopic or atomic systems has found applications in disparate systems: from solid state devices \cite{GiazottoNature2012,Dutta2017} to light harvesting complexes \cite{Caruso2009,Schlawin2013,LeggioEPL2015a,DoyeuxPRA2017}; from ultra-cold atomic systems \cite{Brantut2013,Brunelli2016a,Landig2016} to trapped ions \cite{Barreiro2011,Bermudez2013,Freitas2016}. 
On the other side, coherent manipulation of energy is at the core of quantum thermodynamics, whose aims include the understanding of the emergence of thermodynamic laws from quantum mechanics \cite{Jarzynski1997,Crooks1998,Tasaki2000,Mukamel2003,Jarzynski2004a,CampisiRMP2011,XuerebReview,AndersReview,GooldReview} and the design of thermal machines made with quantum devices \cite{Scovil1959, Geva1992,HePRE2002,QuanPRE2007,LindenPRL2010,Kosloff2014,CorreaSR2014,ZhengPRE2014,Uzdin2015,LeggioPRA2015,LeggioEPL2015b,LeggioPRE2016,DoyeuxPRE2016,AbahEPL2016,NiedenzuNJP2016,KosloffEntropy2017,RouletPRE2017,Reid2017,Scopa2018a,Hewgill2018}.

The starting point for addressing these phenomena is the theory of open quantum systems \cite{Heidelberg1987,Gardiner2004,Breuer2007,rivas2012open}. 
In this framework, the reduced dynamics of the state of the system under scrutiny, when in contact with an environment, is cast in the form of a  master equation after a series of approximations. Under this category fall very common models studied in the literature, such as the spin-boson or the Caldeira-Leggett models \cite{Caldeira1981a,LeggettRMP1987,Caldeira2014b}.

\begin{figure*}[t]
\centering
\includegraphics[width=0.95\textwidth]{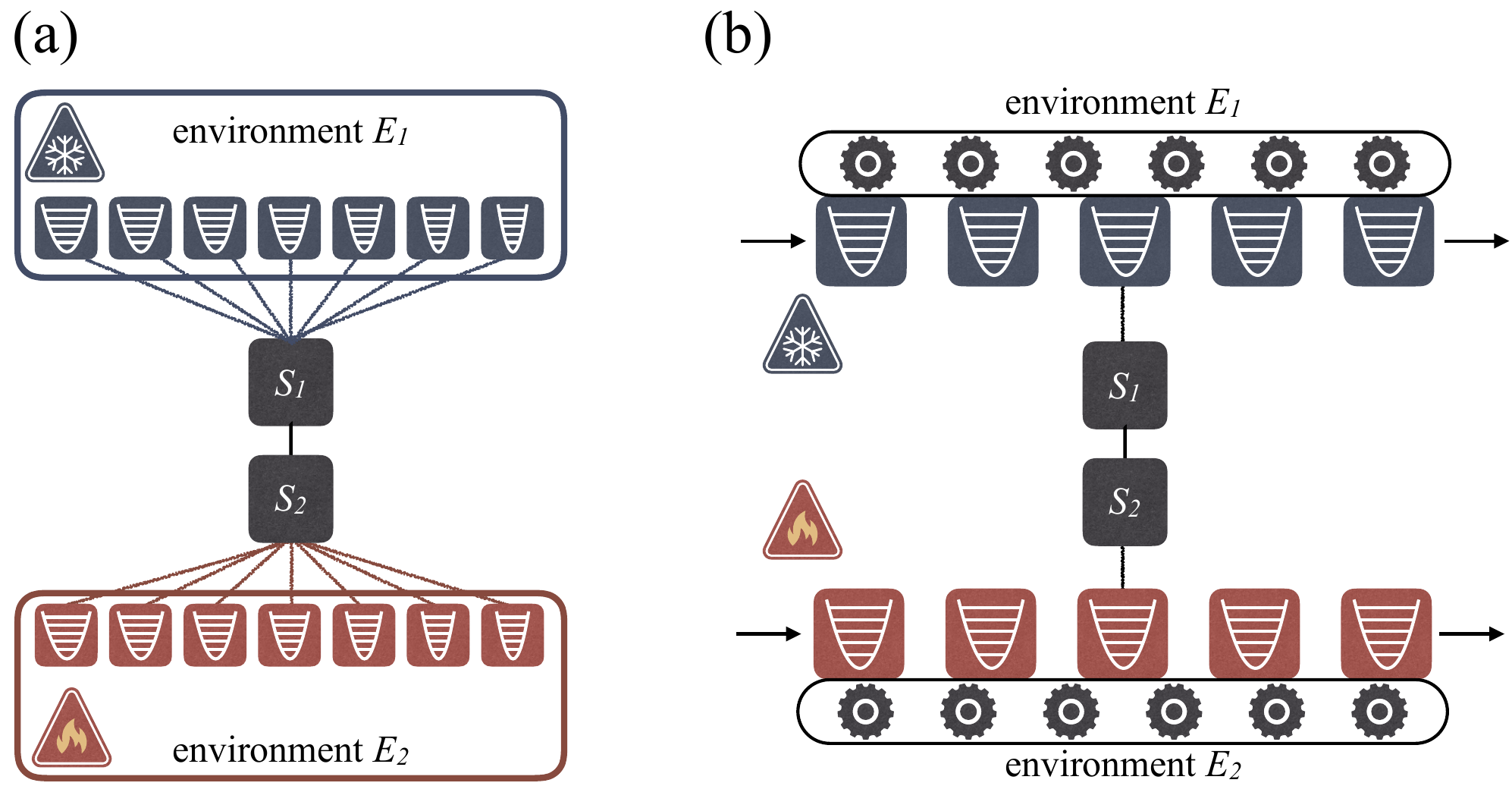}\\
\caption{\label{fig:drawings}
We consider in this paper a system $S$ composed of several sub-systems $S_1, \ldots, S_N$ (in the figure $N = 2$). 
Each sub-system $S_i$ is connected to a local environment $E_i$ prepared in a different temperature $T_i$. 
(a) The standard bosonic heat bath model: the environment is assumed to consist of an ensemble of independent quantum harmonic oscillators with different frequencies in thermal equilibrium and coupled permanently to the system.
(b) In this paper we focus instead on the framework of the repeated interactions method: the environment $E_i$ is divided into a series of ancillas (in this case represented by individual bosonic modes with identical frequencies) which interact with $S_i$ sequentially. This type of method leads to local master equations (LMEs), irrespective of the 
internal interactions between $S_i$ and $S_j$.
}
\end{figure*}

The situation becomes more involved when the system $S$ is made up of several interacting subsystems $S_1, \ldots, S_n$ and each subsystem interacts with a local environment $E_i$, possibly held at different temperatures $T_i$ (see Fig.~\ref{fig:drawings}(a)). 
This type of scenario is the basis for the description of transport properties in quantum systems. 
Microscopic derivations in this case usually lead to global master equations (GMEs),  in which the environment introduces jump operators that allow for  incoherent excitation transfers between the different subsystems \cite{Rivas2010b,Purkayastha2016,Santos2016,Gonzalez2017,HoferNJP2017}. 
However, these derivations are in general quite involved since they require knowledge of the full set of eigenvalues and eigenvectors of the system's Hamiltonian,  something which quickly becomes prohibitive when the number of subsystems increases.
Moreover, depending on the approximations employed, one may also arrive at equations which do not generate  completely positive maps (the so-called Redfield equations \cite{Purkayastha2016}), or equations which contain unphysical heat currents \cite{Wichterich2007}. 
For these reasons, microscopic derivations of master equations for systems connected to multiple environments still continues, nowadays, to be a topic of great interest.

An alternative, more heuristic, approach  consists in deriving a master equation for the individual subsystems, neglecting the interaction with the remaining subsystems. 
The resulting master equation will then contain only local jump operators describing exchanges  between the environment $E_i$ and its corresponding subsystem $S_i$.
Such equations, which we shall henceforth refer to as local master equations (LME) (also frequently called boundary-driven master equations), are typically accurate when the dissipation rates are larger than the interaction between subsystems. 
Due to their computational simplicity, they have been extensively employed over the last two decades in the study of transport in non-equilibrium quantum systems
\cite{Karevski2009,Karevski2013,Asadian2013,Landi2014b,Landi2015a,Schuab2016a,Guimar2016,Cui2015,Prosen2011b,Prosen2011,Prosen2008a,Prosen2012,Prosen2013a,NicacioPRE2015,CampbellSR2016,NicacioPRA2016,Ozgur1,Chanda,PereiraPRE2018}.

It turns out, however, that the nonlocal terms neglected in the LME may still lead to non-thermal steady states~\cite{Cresser1992} and play a significant role if the heat exchanges are small, even for weakly interacting parts.
As a consequence, it has been found that LMEs may lead to apparent  thermodynamic inconsistencies, as pointed out recently by Levy and Kosloff \cite{LevyKosloffEPL2014}.   
They have shown that the LME for two coupled quantum harmonic oscillators (QHO) may predict currents from a cold to a hot thermal reservoir, or the existence of currents even in the absence of a temperature gradient. 
Moreover, both may  occur even in the limit of weakly coupled oscillators.
The origin of these effects, as shown in  \cite{LevyKosloffEPL2014}, lies in the fact that the heat fluxes become of the same order of magnitude as the neglected nonlocal terms.  
They thus suggest to use the GME  to fix such thermodynamic anomalies. 
These findings generated a considerable stream of investigations on the comparison between global and local master equations \cite{Rivas2010b,CorreaPRE2013,Guimar2016,Volovich, Gonzalez2017,HoferNJP2017,Stockburger2017,Decordi2017,MitchisonNJP2018,naseem2018thermodynamic,shammah2018open} as well as clever possible alternatives~\cite{Katz2016,SeahPRE2018}.  

Recent results indicate, however, that it is possible to construct a consistent thermodynamic framework for LMEs, resolving these seeming contradictions. 
First and foremost, it is important to mention that, unlike Redfield equations, LMEs are in Lindblad form and  therefore generate completely positive trace preserving (CPTP) maps. 
Second, LMEs, being CPTP maps, can be microscopically derived in a controlled way  using the idea of repeated interactions (collisional models)~\cite{ScaraniPRL2002,ZimanPRA2002,Karevski2009,PalmaGiovannetti,Landi2014b,Barra2015,Strasberg2016,Manzano2016}. 
In this approach, each environment is divided into an ensemble of identically prepared auxiliary systems, called units, which interact sequentially with an individual subsystem for a short time $\tau$ (see Fig.~\ref{fig:drawings}(b)).
In the limit $\tau\to 0$ this leads to a LME, irrespective of the internal system interactions. 
Thirdly, Barra recently showed that since the interaction between the auxiliary units and the system is time dependent, there is an inherent external work required for generating  the dissipative evolution described by the LME~\cite{Barra2015}. Hence, it is possible to put forth a consistent thermodynamic framework for the repeated interactions scheme, including definitions of heat, work and entropy production \cite{Barra2015,Strasberg2016}.

The goal of  this paper is to advance further on this reconciliation between  LMEs and thermodynamics.
Using the techniques of eigenoperators, we show that while LMEs  satisfy local detailed balance, in general global detailed balance is broken. 
Moreover, we find that there is a fundamental work cost  associated with this breaking of global detailed balance.
By taking this exact work cost into account, we then show that  LMEs become fully reconciled with thermodynamics. 
Quite surprisingly, the discrete nature of the repeated interactions method  allows us to establish a direct mapping between non-equilibrium steady-states and limit-cycles of quantum heat engines. 
Hence, we find that the anomalous situation in which heat flows from cold to hot baths corresponds simply to a refrigerator, requiring positive injection of work to operate.

As our main application, we focus on the case of a  system of two coupled  quantum harmonic oscillators (QHO), which has been the subject of considerable interest due to potential applications in trapped ions \cite{Barreiro2011,Bermudez2013}, optomechanical systems \cite{Ockeloen-Korppi2018,Brunelli2016a} and ultra-cold atoms \cite{Brunelli2016a,Landig2016}. 
Several interesting features of this model are studied in detail. 
First we show how to engineer the system to behave as a refrigerator, a heat engine or an accelerator. 
We study the efficiency and/or coefficient of performance for this model and show that they reach the optimal value of the Otto cycle in the case where  the oscillators interact without counter-rotating terms in the Hamiltonian: including counter-rotating terms only degrades the operation of the system.
We also discuss how to reconcile these results with a theorem  by Martinez and Paz~\cite{MartinezPaz}, which states that no autonomous refrigeration is possible with equilibrium harmonic oscillators linearly coupled. 

Let us mention that, given a physical system, the actual choice of whether to describe it using an LME or a GME approach depends ultimately on the system itself. Here we stress that the LME approach {\it per se} does not give rise to thermodynamic inconsistencies, provided the mechanism for its emergence is fully taken into account. 

In order to clearly present the motivation behind our study and the main features of our approach, we begin in Sec.~\ref{sec:ss} with the simplest example of a  LME, consisting of two coupled QHOs. We first review the seeming thermodynamic inconsistencies that may stem from this model and then go on to show how the results obtained using the method of repeated interactions can be used to completely remove them. 
Then, in Sec.~\ref{sec:RI} we carry out our main theoretical development, showing how the  eigenoperator technique may be used to directly access thermodynamic quantities and describe the possible breaking of global detailed balance. 
Finally, in Sec.~\ref{sec:qho} we return to the oscillator model and  show how the full machinery of our approach may be employed to study the interplay between different types of interactions between the two oscillators. Finally, in Sec.~\ref{sec:conclusions} we summarise our results and conclude.
Our paper also contains three appendices with various details of the derivations of our main results. In Appendix~\ref{sec:lyap} we review the technique of Lyapunov equations for dealing with Gaussian continuous variable systems. In Appendix~\ref{sec:details} we provide additional mathematical details on the developments of Sec.~\ref{sec:RI}. In Appendix~\ref{sec:prod} we compute the entropy production showing that it is always positive.

%
%
\section{\label{sec:ss}Steady-state of two harmonic oscillators subject to  a local master equation}
%
%

In order to motivate our study and provide an intuitive summary of our main results, we begin by discussing the model studied by Levy and Kosloff \cite{LevyKosloffEPL2014}, consisting of  two coupled harmonic oscillators, $S_1$ and $S_2$, each coupled to  its own local heat bath via a LME (see Fig.~\ref{fig:drawings}(a)). 
The Hamiltonian is taken to be 
\begin{equation}\label{eq:H2}
H_S = H_{1}+H_{2} + H_{I},
\end{equation}
where
\begin{equation}\label{HSi}
H_{i} = \frac{1}{2}(p_i^2 +  \omega_i^2 x_i^2) = \omega_i (a_i^\dagger a_i + \nicefrac{1}{2}),
\end{equation}
and 
\begin{equation}\label{HI}
H_{I} = \epsilon (a^\dagger_1a_2+a_1a^\dagger_2).
\end{equation}
Here $a_i$ are the annihilation operators of each harmonic oscillator and $(x_i, p_i)$ the corresponding position and momentum ($\hbar = k_B = 1$). 
Both oscillators are assumed to have unit mass, but  different frequencies $\omega_i$. 
The interaction~(\ref{HI}) is taken to be of the typical tight-binding form, which conserves the number of quanta in the case of equal frequencies, $\omega_1=\omega_2$. 
More general interactions will be discussed in Sec.~\ref{sec:qho}

The system is also subject to two baths $E_1$ and $E_2$, which we assume can be modeled by LMEs acting only on $S_1$ and $S_2$. 
Hence, we take the system to evolve according to  the Lindblad master equation
\begin{equation}
\label{eq:MElocaloscillators}
\frac{\ud \rho_S}{\ud t} = -i [H_S, \rho_S] + D_1(\rho_S) + D_2(\rho_S),
\end{equation}
where 
\begin{equation}\label{qhos_Di}
D_i(\rho_S) = \gamma(n_i+1)\mathcal L[a_i,\rho]+\gamma n_i\mathcal L[a^\dagger_i,\rho].
\end{equation}
Here $\gamma$ is the dissipation rate, $n_i$ is the average excitation number for the Bose-Einstein distribution,
\begin{equation}\label{BoseEinstein}
n_i=\frac{1}{e^{\beta_i\omega_i}-1},
\end{equation}
with inverse temperature $\beta_i=(T_i)^{-1}$ and 
\begin{equation}\label{lindblad_dissipator}
\mathcal L[A,\rho]=A\rho A^\dagger - \frac{1}{2}\{A^\dagger A  ,\rho\},
\end{equation} 
is the dissipator in Lindblad form.
Henceforth, we always assume that $T_1 < T_2$  (see Fig.~\ref{fig:drawings}(a)).
Below, it will also be useful to keep in mind that $n_i$ is a monotonically increasing function of $T_i/\omega_i$.

\subsection{\label{sec:LME}Predictions for the steady-state}

Due to the form of the Hamiltonian and the jump operators, the steady-state of the master equation~(\ref{eq:MElocaloscillators}) will be Gaussian
and can therefore be easily found using the standard Lyapunov equation technique.
A review of this method is given for completeness in Appendix~\ref{sec:lyap}. 
Here we only focus on the main results, which can be summarized by the following expectation values: 
\begin{IEEEeqnarray}{rCl}
\label{2QHOs_occ1}
\langle a_1^\dagger a_1 \rangle &=& n_1 + \frac{2\epsilon^2}{\Delta^2} (n_2 - n_1),
\\[0.2cm]
\label{2QHOs_occ2}
\langle a_2^\dagger a_2 \rangle &=& n_2 - \frac{2\epsilon^2}{\Delta^2} (n_2 - n_1),
\\[0.2cm]
\label{2QHOs_occ3}
\langle a_2^\dagger a_1  \rangle &=& \frac{\epsilon}{\Delta^2}(\omega_1 - \omega_2 + i \gamma)(n_1 - n_2).
\end{IEEEeqnarray}
where $\Delta^2 = \gamma^2 + 4 \epsilon^2 + (\omega_1 - \omega_2)^2$. 
Using Eq.~(\ref{HSi}) we then find that the energy of  oscillator 1 will be 
\begin{equation}
\label{eq:deltae}
\langle H_1 \rangle =\langle H_1 \rangle_\text{th} + 
\frac{2 \omega_1 \epsilon ^2}{\Delta^2} (n_2-n_1),
\end{equation}
where $\langle H_1 \rangle_\text{th} = \omega_1 (n_1+ \nicefrac{1}{2})$ is the energy the oscillator would have if it were in thermal equilibrium with its corresponding bath.
An equivalent expression can be obtained for the second oscillator replacing $1\leftrightarrow 2$. 
Similarly, the heat  flux from the first oscillator to the second will be 
\begin{equation}\label{Q21}
\dot Q_{1\to 2} = -i \left\langle [H_S,H_1]\right\rangle= -i\omega_1\epsilon\langle a^\dagger_1a_2-a_1a^\dagger_2\rangle=\frac{2\gamma\epsilon^2\omega_1}{\Delta^2}(n_2-n_1),
\end{equation}
which is consistent with the expression obtained in Ref.~\cite{LevyKosloffEPL2014}.

The relevant point to highlight from these results is that the sign of the $\epsilon^2$-terms in Eqs.~(\ref{eq:deltae}) and (\ref{Q21}) will depend on the relative magnitudes of $n_1$ and $n_2$ and {\it not} of $T_1$ and $T_2$. 
This simple  fact, when combined with the definition of $n_i$ in Eq.~(\ref{BoseEinstein}), introduces several properties which, at first sight, seem to violate the second law of thermodynamics. 
First, it implies that oscillator 1 can be cooled to a temperature smaller than the temperature $T_1$ of the coldest reservoir, in the sense that it could reach a state with lower energy than that of the thermal state it would reach if it were in contact only with the coldest reservoir. 
Second, one may obtain a non-zero heat current even if both baths are at the same temperature,  $T_1=T_2$. 
And third, if $T_1 < T_2$ but we happen to satisfy 
\begin{equation}\label{refrigerator_condition}
n_1 > n_2 \quad \leftrightarrow \quad \frac{T_1}{\omega_1}>\frac{T_2}{\omega_2},
\end{equation}
then there will be a current flowing from the cold to the hot bath ($\dot Q_{1\to 2}<0$).

As pointed out in Ref.~\cite{LevyKosloffEPL2014}, these results seem to violate the second law of thermodynamics, a fact which the authors attribute to the inadequacy of local master equations to describe transport processes. The result of the cooling of the first oscillator below the temperature of the coldest environment also seems  to suggest that this master equation could be used to design Gaussian (quadratic) absorption refrigerators, which would contradict a general theorem proved by Martinez and Paz~\cite{MartinezPaz}. 

\begin{center}
\begin{table*}
\caption{\label{tab}
The three modes of operation of the steady-state of the two oscillators, always assuming $E_1$ is the cold reservoir ($T_1 < T_2$). 
In this paper positive heat or work always means energy entering the system. 
So $\dot Q_i > 0$ means energy entered $S$ through $E_i$ and $\dot W_{\rm ext}>0$ means work was performed on the system by an external agent.
}
\begin{tabularx}{\textwidth}{bmsssc}
\hline\\[-0.1cm]
Mode of operation &  &  $\dot Q_1$ & $\dot Q_2$ & $\dot W$ & Operation regime	\\[0.2cm]
\hline\\[-0.1cm]
{\bf Refrigerator} & &  $\dot Q_1>0$ 	&  $\dot Q_2 <0$ & $\dot W >0$  & $\omega_1 < \omega_2 T_1/T_2$ \\
Consumes work to make heat flow from cold to hot. & &&  & & \\[0.2cm]
{\bf Engine}	   &&   $\dot Q_1<0$ 	&  $\dot Q_2 >0$ & $\dot W <0$  & $ \omega_2 T_1/T_2 <\omega_1 < \omega_2$ \\
Uses hot bath to produce useful work, dumping the remainder in the cold bath. & & & & & \\
\\

{\bf Accelerator (oven)}	  &  &  $\dot Q_1<0$ 	&  $\dot Q_2 >0$ & $\dot W >0$  & $\omega_1 > \omega_2$ \\
Consumes work to heat the cold reservoir more than it would with spontaneous thermal conduction.\\[0.2cm]
\hline
\end{tabularx}
\end{table*}
\end{center}

\subsection{Resolving the thermodynamic inconsistencies}

Using Eq.~(\ref{eq:MElocaloscillators}) one finds that the total rate of change of the system Hamiltonian~(\ref{eq:H2}) is given by 
\begin{IEEEeqnarray}{rCl}
\frac{\ud \langle H_S \rangle}{\ud t} &=& \gamma (n_1 - \langle a_1^\dagger a_1 \rangle ) + \gamma (n_2 - \langle a_2^\dagger a_2 \rangle) 
\nonumber\\[0.2cm]
&&- \gamma \epsilon \langle a_1^\dagger a_2 + a_2^\dagger a_1 \rangle.
\label{dHSdt}
\end{IEEEeqnarray}
However, this change in energy cannot be associated only with heat flowing to the reservoirs \cite{Barra2015}. 
Instead, there is an associated work cost which, as we show below, is related to  the breaking of global detailed balance.
Quoting the results that will be derived in Sec.~\ref{sec:RI}, we now discuss how 
one can remove all the inconsistencies by including this work cost.

In particular, we find that the correct splitting of Eq.~(\ref{dHSdt})  is of the form 
\begin{equation}
\frac{\ud \langle H_S \rangle}{\ud t} = \dot{Q}_1 + \dot{Q}_2 + \dot{W}_{\rm ext},
\end{equation}
where
\begin{equation} \label{heat_cost_2QHOs}
\dot{Q}_i ={\rm tr} \left [D_i(\rho_S) H_i\right ]= \gamma \omega_i (n_i - \langle a_i^\dagger a_i \rangle),
\end{equation}
is the heat rate  flowing from each reservoir to the system   (see Eq.~(\ref{heat_rate}) for the general expression) and 
\begin{equation}\label{work_cost_2QHOs}
\dot{W}_{\rm ext}= {\rm tr} \left\{\left [D_1(\rho_S)+D_2(\rho_S)\right ] H_I\right\}=- \gamma \epsilon \langle a_1^\dagger a_2 + a_2^\dagger a_1 \rangle .
\end{equation}
is the required work rate (see Eq.~(\ref{work_rate}) for the general expression).
From Eqs.~(\ref{2QHOs_occ1})-(\ref{2QHOs_occ3}) we then find that in the steady-state
\begin{IEEEeqnarray}{rCl}
\label{2qhos_heat_rate_1_explicit}
\dot Q_1 &=& \frac{2\gamma\epsilon^2}{\Delta^2}\omega_1(n_1-n_2),
\\[0.2cm]
\label{2qhos_heat_rate_2_explicit}
\dot Q_2 &=& -\frac{2\gamma\epsilon^2}{\Delta^2}\omega_2(n_1-n_2),
\\[0.2cm]
\dot{W}_{\rm ext} &=& -\frac{2\gamma\epsilon^2}{\Delta^2} (\omega_1 - \omega_2)(n_1 - n_2).
\label{2qhos_work_rate_explicit}
\end{IEEEeqnarray}

These are very important results, which we shall now discuss in depth. 
But before doing so, some consistency checks are in order. 
First, $\dot{W}_{\rm ext}$ tends to zero as $\epsilon\to 0$, in which case global detailed balance is recovered. 
Second, the work required is proportional to $\gamma\epsilon^2$.
As discussed in Ref.~\cite{Gonzalez2017}, the local master equation can be shown to be correct for small interactions $\epsilon$ up to the same order (see comment in Sec. II~D in that paper). 
Third, in the steady-state $\ud \langle H_S \rangle/\ud t = 0$ so that we should have
\begin{equation}
\dot{W}_{\rm ext} = - \dot{Q}_1 - \dot{Q}_2,
\end{equation}
as can indeed be easily verified.

\begin{figure*}
\centering
\includegraphics[width=0.19\textwidth]{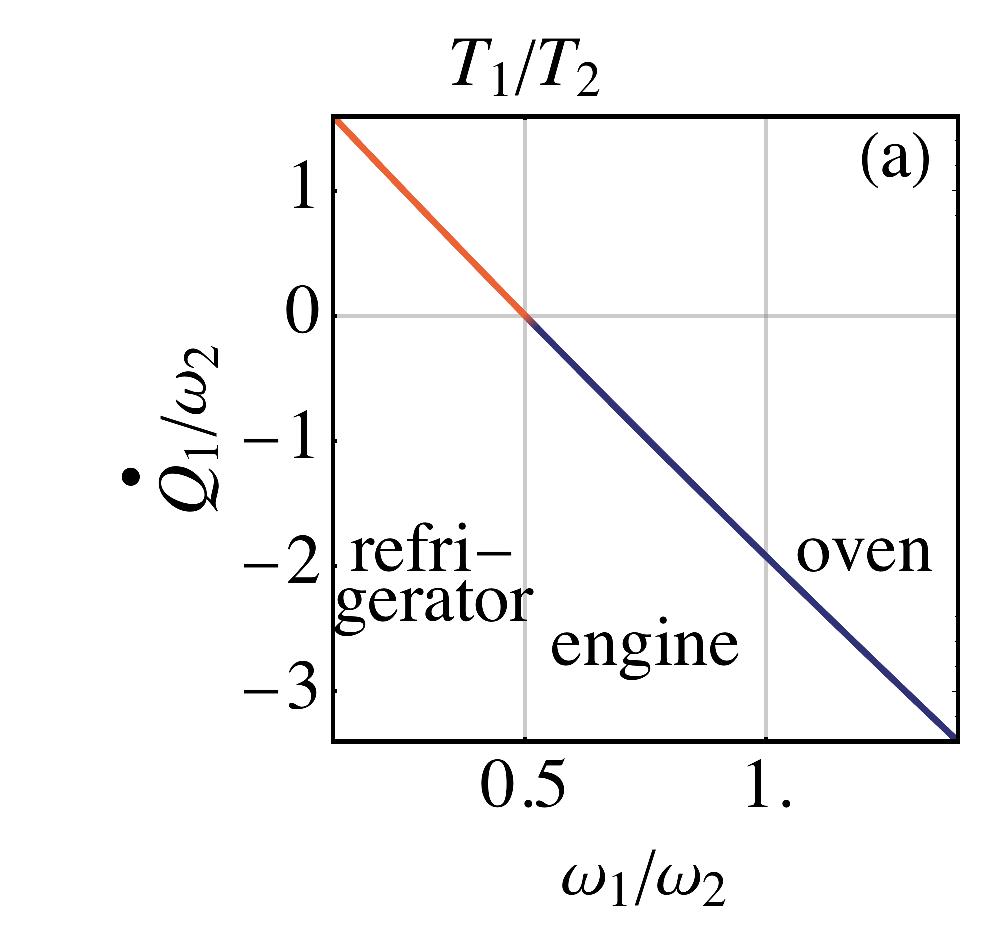}\;
\includegraphics[width=0.19\textwidth]{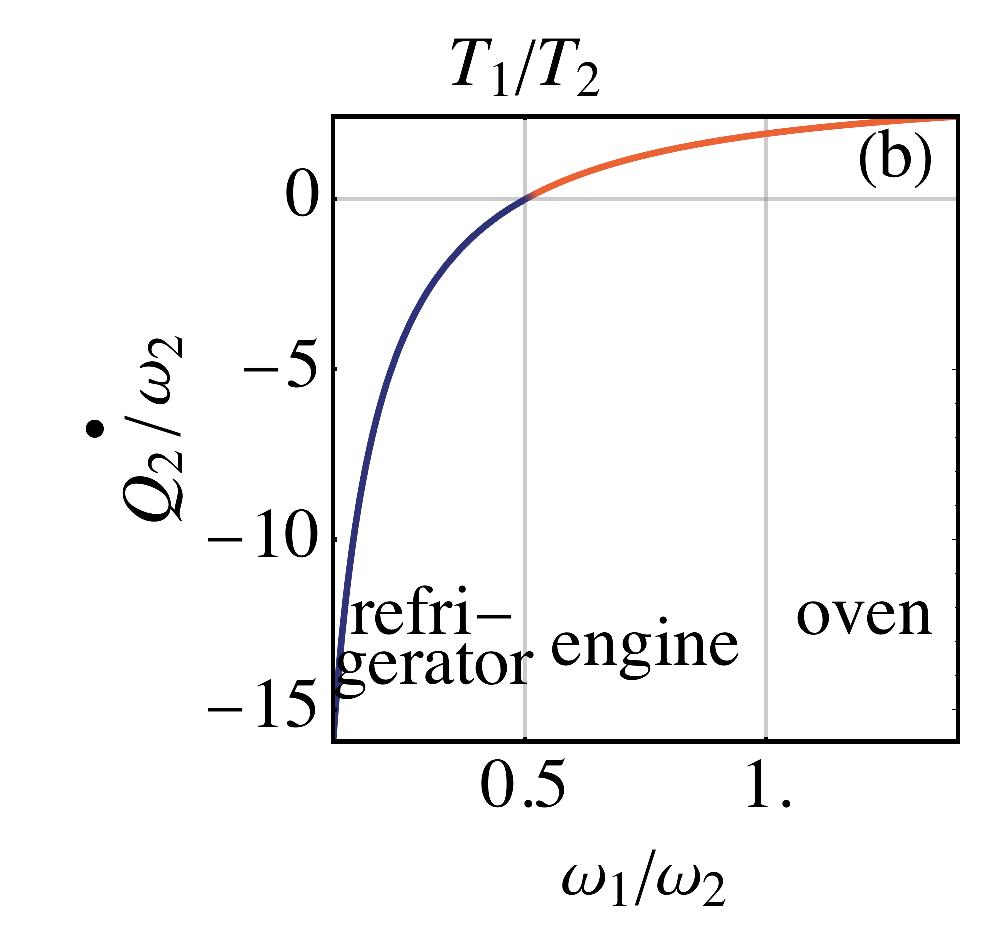}\;
\includegraphics[width=0.19\textwidth]{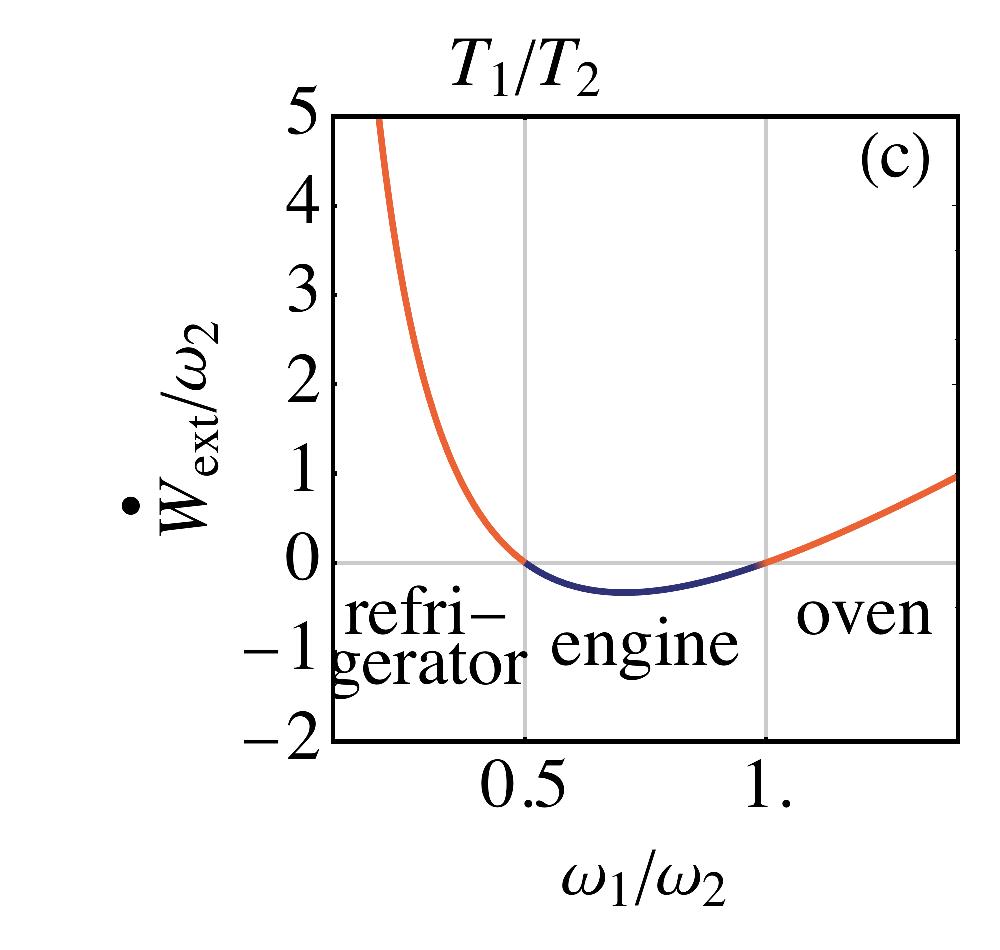}\;
\includegraphics[width=0.19\textwidth]{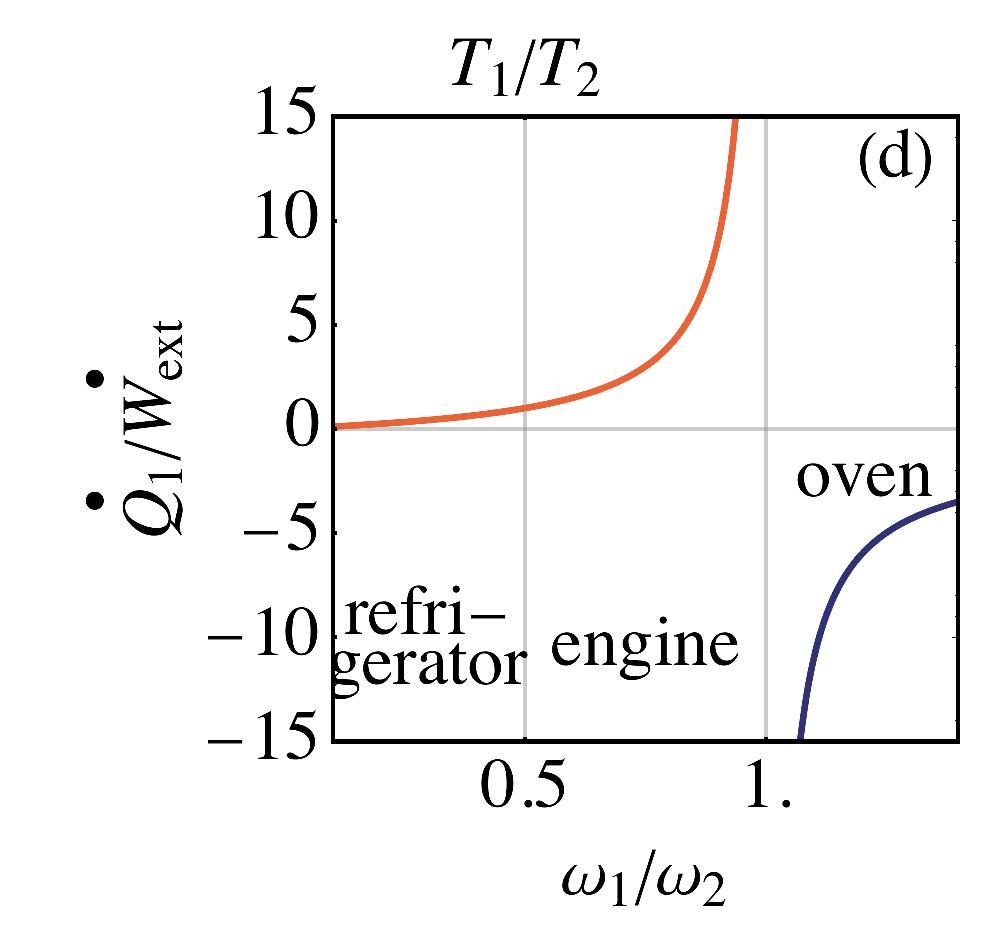}\;
\includegraphics[width=0.19\textwidth]{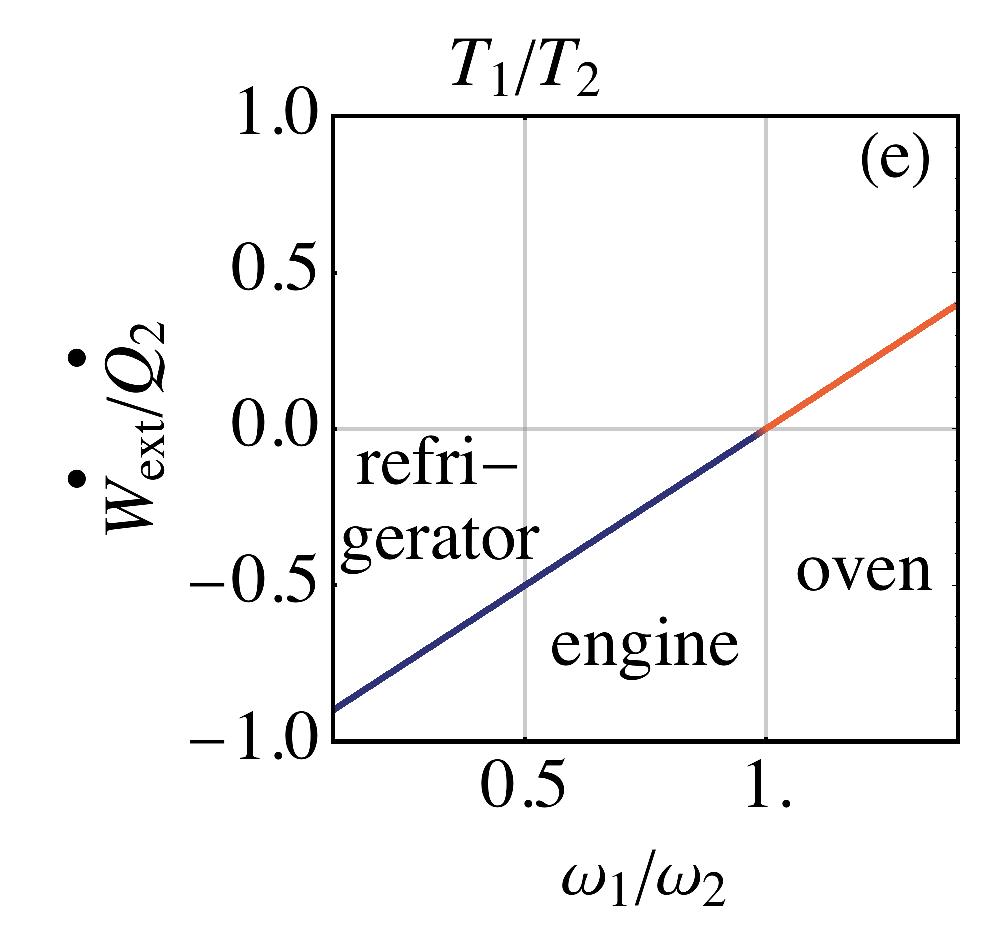}
\caption{\label{fig:2QHOs}
The three regimes of operation in Table~\ref{tab}, for the steady-state of two harmonic oscillators. 
(a)-(c) $\dot Q_1$, $\dot Q_2$ and $\dot W_{\rm ext}$,  Eqs.~(\ref{2qhos_heat_rate_1_explicit})-(\ref{2qhos_work_rate_explicit}), as a function of $\omega_1/\omega_2$. 
(d) $\dot Q_1/\dot W_{\rm ext}$, which follows precisely the Otto value $\omega_1/(\omega_2 - \omega_1)$.  
(e) $\dot W_{\rm ext}/\dot Q_2$ which also follows the Otto value $\omega_1/\omega_2  - 1$. 
All curves are plotted for $\gamma \epsilon^2/\Delta^2 = \omega_2$, assuming $T_1/T_2 = 1/2$. 
The vertical lines represent $\omega_1/\omega_2 =T_1/T_2$ and $\omega_1=\omega_2$. 
}
\end{figure*}

We now show that, once this work cost is correctly attributed, the two-oscillator system at steady-state functions precisely as a heat engine, with two reservoirs and a work source. 
Indeed, we find from Eqs.~(\ref{2qhos_heat_rate_1_explicit})-(\ref{2qhos_work_rate_explicit}) that the system can be tuned to function as a refrigerator, a thermal engine or an accelerator (oven). 
The description of each mode of operation and what they mean in terms of $\dot Q_1$, $\dot Q_2$ and $\dot W$ is presented in Table~\ref{tab}.
We also show in Fig.~\ref{fig:2QHOs} a plot of these 3 quantities illustrating the typical operation regimes. 

We begin with the case of a refrigerator, $\dot Q_1 > 0$. 
According to Eq.~(\ref{2qhos_heat_rate_1_explicit}), this requires $n_1 > n_2$ which is tantamount to the condition~(\ref{refrigerator_condition}).
But we now see that when this condition is satisfied, one also has  $\dot W_{\rm ext} > 0$.
Thus,  for the two  oscillators to act as a refrigerator, they must absorb work from an external agent.
Hence, we see that there are no violations of the second law. 
Instead, by taking into account this work cost, one recovers a fully consistent thermodynamic description.
The coefficient of performance (COP) of the refrigerator is the ratio of the heat extracted from the cold environment divided by the work invested:
\begin{equation}
\label{eq:COPoscillators}
{\rm COP}=\frac{\dot Q_1}{\dot W_{\rm ext}} = \frac{\omega_1}{\omega_2-\omega_1}
\end{equation}
which coincides with the COP of the Otto cycle. 
As will become clear in Sec.~\ref{sec:RI}, this connection between a steady-state and the limit-cycle of a heat engine, can actually be traced back to the basic idea of the repeated interactions method,  which models the steady-state as a sequence of (extremely short) strokes, each having an associated work rate and heat rate. 
Notwithstanding, we should also mention that an Otto COP is a special consequence of this type of interaction. 
As will be shown in Sec.~\ref{sec:qho},  the introduction of counter-rotating terms only reduces the COP  below the Otto value. 

Increasing $\omega_1$ in the interval 
\begin{equation}
\frac{T_1}{T_2} \omega_2 < \omega_1 < \omega_2 ,
\end{equation}
we find that the two-oscillator system in its steady-state functions as a thermal machine,  extracting heat from the hot bath, producing work and dumping heat in the cold bath. 
In this regime the machine's efficiency will be:
\begin{equation}
\eta=\frac{|\dot W_{\rm ext}|}{\dot Q_2} = 1-\frac{\omega_1}{\omega_2}
\end{equation}
which is again the Otto efficiency for a thermal machine. 
Finally, for  $\omega_2<\omega_1$ we necessarily have $n_1<n_2$ and therefore  $\dot Q_1<0$, $\dot Q_2>0$ and $\dot W_{\rm ext}>0$. Thus the device transports heat from the hot to the cold environment and at the same time transforms work into heat which is damped in the cold bath: $|\dot Q_1|=\dot W_{\rm ext}+\dot Q_2>\dot Q_2$. Hence, it functions as an accelerator (i.e. an oven) \cite{Buffoni2018}, heating the cold reservoir faster than it would with spontaneous thermal conduction.
 
 Notice that at the special value $\omega_1/\omega_2 = T_1/T_2$, the system operates as a thermal machine with the Carnot efficiency and zero power.
 
We conclude this section  by emphasising that within the repeated interactions framework one is able to resolve all of the seeming thermodynamic inconsistencies of local master equations. The zeroth law does not apply since the system is externally driven.
The first law is satisfied, as we  accounted for the energy balance of the whole system plus environment, including the required work cost. 
The second law is satisfied, as refrigeration is not spontaneous, but accompanied by external work. 
And finally, the LMEs  do not violate the theorem by Martinez and Paz~\cite{MartinezPaz}, which states that no autonomous refrigeration is possible with equilibrium harmonic oscillators. 
In fact, in our framework, there is also work associated with the environment-system interaction so the machine is not autonomous. 

We now move on to the mathematical development of LMEs using eigenoperators and the repeated interactions.

%
%
\section{\label{sec:RI}The method of repeated interactions}
%
%

We now turn to the  microscopic derivation of LMEs using the method of repeated interactions \cite{ScaraniPRL2002,ZimanPRA2002,Karevski2009,PalmaGiovannetti,CiccarelloPRA2013,LorenzoPRA2015,Landi2014b,LorenzoPRL2015,Barra2015,Strasberg2016,Manzano2016,Ciccarello2017,PezzuttoNJP2016,pezzutto2018out,cusumano2018entropy,CampbellPRA2018,GrossQST2018}, which will form the basis for our thermodynamic description. 
The basic idea behind the method of repeated interactions is to model the  environment  as a set of units, each prepared in a thermal state. 
At time $t = 0$ the  system $S$ is allowed to interact for a certain time $\tau$ with one such unit, which we generically refer to as $E$.
After this time the unit is discarded and a fresh new one is introduced, again in the same initial state. 
The process is then repeated sequentially. 
For $\tau \to 0$ this method  generates a continuous time description which can be modelled by a Lindblad master equation that is, in general, local. We emphasise that a larger class of non-Markovian master equations including the global master equation can be engineered in a similar way~\cite{PalmaGiovannetti} and realised experimentally as proposed in Ref.~\cite{JinPRA2015}.

We consider here this method implemented in the general scenario of Fig.~\ref{fig:drawings}(b) and derive the results of the previous section as particular cases.
Our  system $S$ is therefore assumed to be composed of $N$ subsystems $S_1, \ldots, S_N$, where each $S_i$ interacts  with its own environment $E_i$. 
The Hamiltonian of the system is taken to be
\begin{equation}\label{HS}
H_S = \sum\limits_{i=1}^N H_{S_i} + H_I,
\end{equation}
where $H_{S_i}$ is the Hamiltonian of subsystem $S_i$ and $H_I$ summarises all interactions between subsystems. 
Moreover, the total Hamiltonian comprising the system, the environments and all interactions, is taken as 
\begin{equation}\label{Htot}
H_\text{tot} = H_S + \sum\limits_{i=1}^N H_{E_i} + \frac{1}{\sqrt{\tau}}\sum\limits_{i=1}^N V_i,
\end{equation}
where $H_{E_i}$ is the Hamiltonian of environment $E_i$ and $V_i$ is the interaction between $S_i$ and $E_i$. 
We also follow the customary approach of scaling the $V_i$ by the interaction time $\tau$, which is convenient, although not necessary, for taking the continuous time limit \cite{Landi2014b,Barra2015,Strasberg2016}.
Following the usual repeated interactions approach, the reduced density matrix of the system $\rho_S(t)$ will then evolve according to the  map 
\begin{equation}\label{strobo}
\rho_S((n+1)\tau) = \tr_E \bigg\{ e^{-i \tau H_\text{tot}} \; \rho_S(n\tau) \rho_E \; e^{i \tau H_\text{tot}}\bigg\},
\end{equation}
where 
\begin{equation}\label{rhoE}
\rho_E = \bigotimes\limits_{i=1}^N \frac{e^{-\beta_i H_{E_i}}}{\tr e^{-\beta_i H_{E_i}}},
\end{equation}
is the thermal state of the environments. 

Expanding Eq.~(\ref{strobo}) in a power series in $\tau$ one then finds, up to first order, the LME,
\begin{equation}\label{M}
\frac{\ud \rho_S}{\ud t} = -i [H_S, \rho_S] + \sum\limits_{i=1}^N D_i (\rho_S),
\end{equation}
where $H_S$ is given in~(\ref{HS}) and 
\begin{equation}\label{Di}
D_i(\rho_S) = - \frac{1}{2} \tr_{E_i} [V_i, [V_i, \rho_S \rho_{E_i}]].
\end{equation}
The interesting aspect of this approach is that the structure of the dissipative terms are completely independent on the choice of $H_S$.
Or, more specifically, on the interaction terms $H_I$. 
This is a consequence of the short interaction times between the system and environment which do not allow for the information of each $E_i$ to scramble towards different $S_j$ \cite{Campisi2017a,Garttner2017}.

\subsection{Local vs. Global detailed balance}

Up to now the discussion makes no reference  to the structure of the system-environment interactions $V_i$. 
A particularly interesting situation, which is in practice the most widely studied case, is when the $V_i$ satisfy local detailed balance with respect to each subsystem, but do not necessarily satisfy global detailed balance due the  system-system interactions $H_I$.
To make this argument more precise we introduce the idea of eigenoperators. 

By using the method of eigenoperators, we can cast the conditions of detailed balance in terms solely of the algebra of operators. This therefore allows us to introduce the notion of local vs. global detailed balance, as referring to the local or global Hamiltonian of the system. We note that here this terminology is being used in a slightly different context as, for instance, Ref.~\cite{Barra2015}. There the term is employed in the standard sense of ``detailed balance"~\cite{KatzPRB1983}.

Let $H$ be an arbitrary Hamiltonian. An operator $A$ is called an eigenoperator of $H$ if it satisfies $[H,A] = - \omega A$, for some  $\omega>0$ (usually referred to as a Bohr, or transition, frequency). 
Due to this algebra, $A$ and $A^\dagger$ function as lowering and raising operators for the $H$, causing transitions between energy levels $E$ and $E'$ separated by an energy $\omega$.

Returning to our problem, we shall  assume that the $V_i$ have the form 
\begin{equation}\label{Vi}
V_i = \sum\limits_k g_{i,k} (L_{i,k}^\dagger A_{i,k} + L_{i,k} A_{i,k}^\dagger),
\end{equation}
where $g_{i,k}$ are constants, while $L_{i,k}$ and $A_{i,k}$ are eigenoperators of $H_{S_i}$ and $H_{E_i}$ respectively. 
That is:
\begin{IEEEeqnarray}{rCl}
\label{eigenop_sys}
[H_{S_i}, L_{i,k}] = - \omega_{i,k} L_{i,k}, 
\\[0.2cm]
\label{eigenop_bath}
[H_{E_i}, A_{i,k}] = - \omega_{i,k} A_{i,k},
\end{IEEEeqnarray}
where, for each $S_i$,  $\{\omega_{i,k}\}$ represents the set of transition frequencies of the Hamiltonian $H_{S_i}$.
The rationale behind these expressions is the following.
For each subsystem $S_i$, the set of Bohr frequencies $\{\omega_{i,k}\}$ can be viewed as the set of  transitions which are {\it activated} in that subsystem, due to the contact with its local bath.
For instance, in the case of a harmonic oscillator, a dissipator such as~(\ref{qhos_Di}) is characterized by an eigenoperator $a$, which only induces transitions between neighboring levels. Hence, the only Bohr frequency would be $\omega$ (the natural frequency of the oscillator). 
Similarly, a system-environment interaction containing $a^2$ would induce transitions with Bohr frequencies $2\omega$.

In an interaction such as~(\ref{Vi}), however, there is also the additional assumption that whenever there is a transition of $+\omega_{i,k}$ in the system, the corresponding transition in $E_i$ occurs with energy $-\omega_{i,k}$
Hence, it follows that
\begin{equation}\label{local_db}
[H_{S_i} + H_{E_i}, V_i] = 0,
\end{equation}
meaning that all the energy that leaves system $S_i$ enters bath $E_i$.
This equation, together with the assumption that the bath is thermal, implies local detailed balance. 

However, a similar relation does not hold for the total system Hamiltonian $H_S$, due to the interaction $H_I$ between subsystems. 
The reason is that, in general, $L_{i,k}$ are not eigenoperators of $H_I$. Hence,
\begin{equation}\label{global_db}
[H_S + H_E, V] = [H_I, V] \neq 0,
\end{equation} 
where $H_E = \sum_{i=1}^N H_{E_i}$ and $V = \sum_{i=1}^N V_i$. 
This means that even though detailed balance may hold locally, it is violated globally. 
As will be shown below, taking into account this fact is essential for providing a consistent thermodynamic description of LMEs. 

Substituting Eq.~(\ref{Vi}) in Eq.~(\ref{Di}) one may obtain a more explicit formula for the LME dissipators. 
To do so one notices that since the environments are in equilibrium, it follows that $\langle A_{i,k} A_{i,q} \rangle = 0$ and $\langle A_{i,k}^\dagger A_{i,q} \rangle \propto \delta_{k,q}$. 
Hence, Eq.~(\ref{Di}) reduces to
\begin{equation}\label{Di_explicit}
D_i(\rho_S) = \sum\limits_{k}  \gamma_{i,k}^- \; \mathcal{L}[L_{i,k}, \rho_S] + \gamma_{i,k}^+  \; \mathcal{L}[L_{i,k}^\dagger, \rho_S]
\end{equation}
where $\mathcal{L}[A,\rho]$ is the Lindblad dissipator defined in Eq.~(\ref{lindblad_dissipator}), whereas
\begin{IEEEeqnarray}{rCl}
\gamma_{i,k}^- &=& g_{i,k}^2 \;  \langle A_{i,k} A_{i,k}^\dagger \rangle,
\nonumber \\
\label{gamma_pm}
\\
\gamma_{i,k}^+ &=& g_{i,k}^2 \; \langle A_{i,k}^\dagger A_{i,k} \rangle.
\nonumber
\end{IEEEeqnarray}
Since the $A_{i,k}$ are eigenoperators of the bath which is in equilibrium at inverse temperature $\beta_i$, it follows that 
\begin{equation}
\frac{\gamma_{i,k}^+}{\gamma_{i,k}^-} = e^{-\beta_i \omega_{i,k}},
\end{equation}
which is another manifestation of local detailed balance.

\subsection{Example: two harmonic oscillators}

To show how these results may be applied, we consider again the two harmonic oscillators  $a_1$ and $a_2$ treated in Sec.~\ref{sec:ss}.
We assume that the environmental units $E_1$ and $E_2$ are themselves harmonic oscillators with operators $b_1$ and $b_2$~~\footnote{Note that it is possible to arrive at the same results by assuming the environmental units to be qubits.}, and we take the total Hamiltonian to be
\begin{equation}
H_\text{tot} = H_S + \sum\limits_{i=1,2} \bigg\{ \omega_i b_i^\dagger b_i + g (a_i^\dagger b_i + b_i^\dagger a_i)\bigg\},
\end{equation}
where $H_S$ is given in Eq~(\ref{eq:H2}).
Since the frequency $\omega_1$ of $S_1$ and $E_1$ are the same, the operators $a_1$ and $b_1$ are eigenoperators of $H_1 = \omega_1 a_1^\dagger a_1$ and $H_{E_1} = \omega_1 b_1^\dagger b_1$.
Hence, local detailed balance, Eq.~(\ref{local_db}), is satisfied. 
However, $a_1$ is not an eigenoperator of $H_S$ due to the interaction term $H_I = \epsilon (a_1^\dagger a_2 + a_1 a_2^\dagger)$ [Eq.~(\ref{HI})] so that global detailed balance is in general broken. 

Applying Eq.~(\ref{Di_explicit}) with $L_{i,k} = a_i$ and $A_{i,k} = b_i$ we then immediately find the master equation~(\ref{eq:MElocaloscillators}) with the transition rates $\gamma_{i,k}^\pm$  given by 
\begin{equation}\label{gamma_pm_qho}
\gamma_i^+ = g^2 \langle b_i^\dagger b_i \rangle := \gamma n_i,
\qquad
\gamma_i^- = g^2 \langle b_i b_i^\dagger \rangle = \gamma (n_i+1).
\end{equation}
which are the coefficients in Eq.~(\ref{qhos_Di}). 

\subsection{Thermodynamics of the repeated interactions method}

We now turn to a description of the thermodynamics of the repeated interactions method. 
Since the global S+E interaction is unitary, we may study the changes in the energy of the system and environment individually for a given stroke. 
The heat $\delta Q_i$ exchanged between the system and environment $i$ in one  stroke must then be simply  $\langle H_{E_i} \rangle_0 - \langle H_{E_i} \rangle_\tau$. 
Dividing by $\tau$ and taking the limit $\tau\to 0$ will give the heat current $\dot{Q}_i$ to environment $E_i$. 
The explicit calculation is postponed to Appendix~\ref{sec:details}. 
As a result, we find the surprisingly simple formula:
\begin{equation}\label{heat_rate}
\dot{Q}_i = \sum\limits_k \omega_{i,k} \bigg\{ \gamma_{i,k}^+ \langle L_{i,k} L_{i,k}^\dagger \rangle - \gamma_{i,k}^- \langle L_{i,k}^\dagger L_{i,k} \rangle \bigg\}.
\end{equation}
For example, in the case of the two harmonic oscillators,   Eq.~(\ref{heat_rate}) gives precisely the heat-rate formula~(\ref{heat_cost_2QHOs}) that  was used in Sec.~\ref{sec:ss}.

In addition to the heat rates, however, there will in general also be a work contribution.
The simplest way of seeing this is to note that for each interaction stroke one must turn the system-environment coupling on and off,  so that the total Hamiltonian~(\ref{Htot}) must actually be time-dependent. 
Since the global S-E system is isolated, the work in an individual stroke between times $n\tau$ and $(n+1)\tau$ may be unambiguously defined as 
\begin{equation}\label{work_single_stroke}
\delta W_{\rm ext} = \int\limits_{n\tau}^{(n+1)\tau} \left \langle \frac{\partial H_\text{tot}}{\partial t} \right\rangle \ud t.
\end{equation}
Again, dividing by $\tau$ and taking the limit $\tau\to 0$ yields the work rate $\dot W_{\rm ext}$. 
As shown in Appendix~\ref{sec:details}, carrying out this computation yields
\begin{equation}\label{work_rate}
\dot{W}_{\rm ext} = \frac{1}{2} \sum\limits_{i,k} \bigg\{ \gamma_{i,k}^- \langle L_{i,k}^\dagger F_{i,k} \rangle - \gamma_{i,k}^+ \langle F_{i,k} L_{i,k}^\dagger \rangle \bigg\} + \text{c.c.},
\end{equation}
where $F_{i,k} = [H_I, L_{i,k}]$ and c.c. stands for complex conjugate. 
Hence, we see that the work rate is related precisely to the non-commutativity of the jump operators $L_{i,k}$ with the system-system interaction $H_I$. 
If global detailed balance, Eq.~(\ref{global_db}), is recovered, then the work cost is identically zero. 
Otherwise, whenever global detailed balance is broken, there will be an associated work cost. 
As an example, in the case of the two QHOs, $[H_I, a_1] = - \epsilon a_2$, so that a direct application of Eq.~(\ref{work_rate}) leads to expression~(\ref{work_cost_2QHOs}).

From a practical point of view, Eqs.~(\ref{heat_rate}) and (\ref{work_rate}) are our main results, as they offer general expressions that may be applied over a broad range of situations. These results generalise Barra's findings~\cite{Barra2015} to arbitrary bath and system's structure. In Appendix~\ref{sec:prod}, we calculate the entropy production showing that it is indeed always positive in agreement with the second law.
We now return to the problem of coupled harmonic oscillators and show how they can be applied to the study of more complex situations. 

\begin{figure*}
\centering
\includegraphics[width=0.19\textwidth]{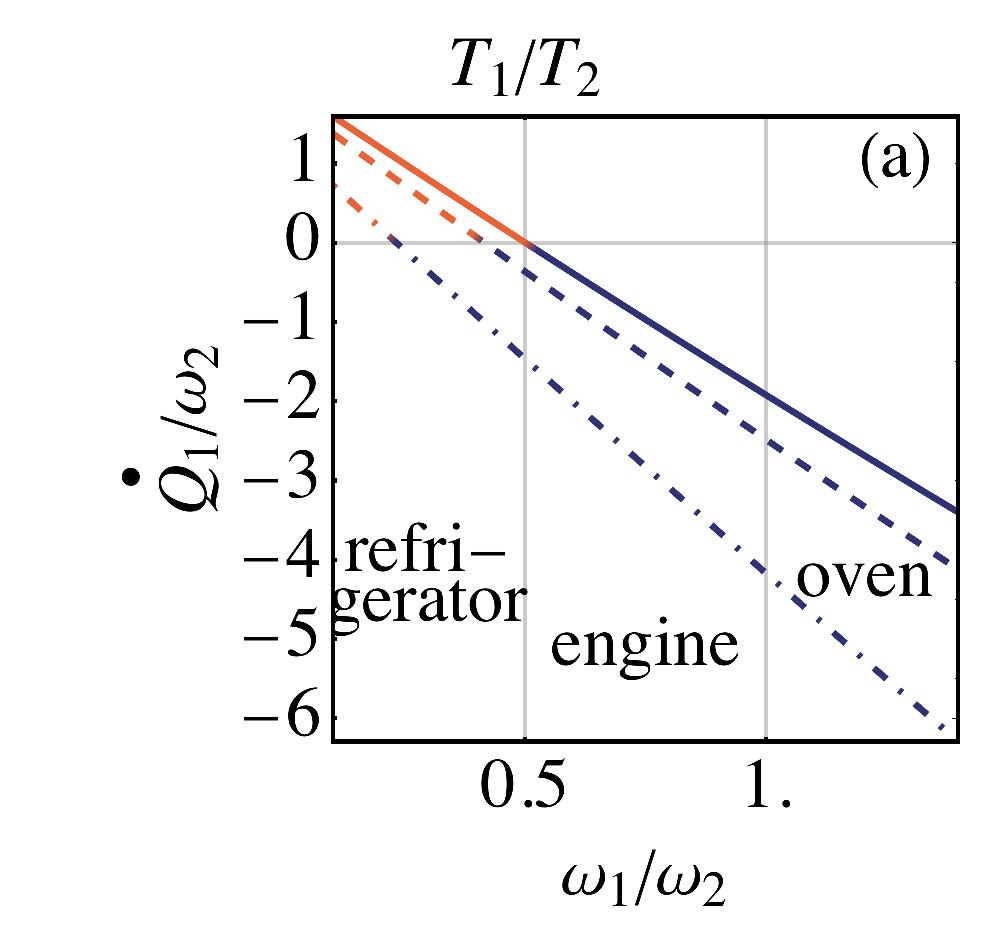}\;
\includegraphics[width=0.19\textwidth]{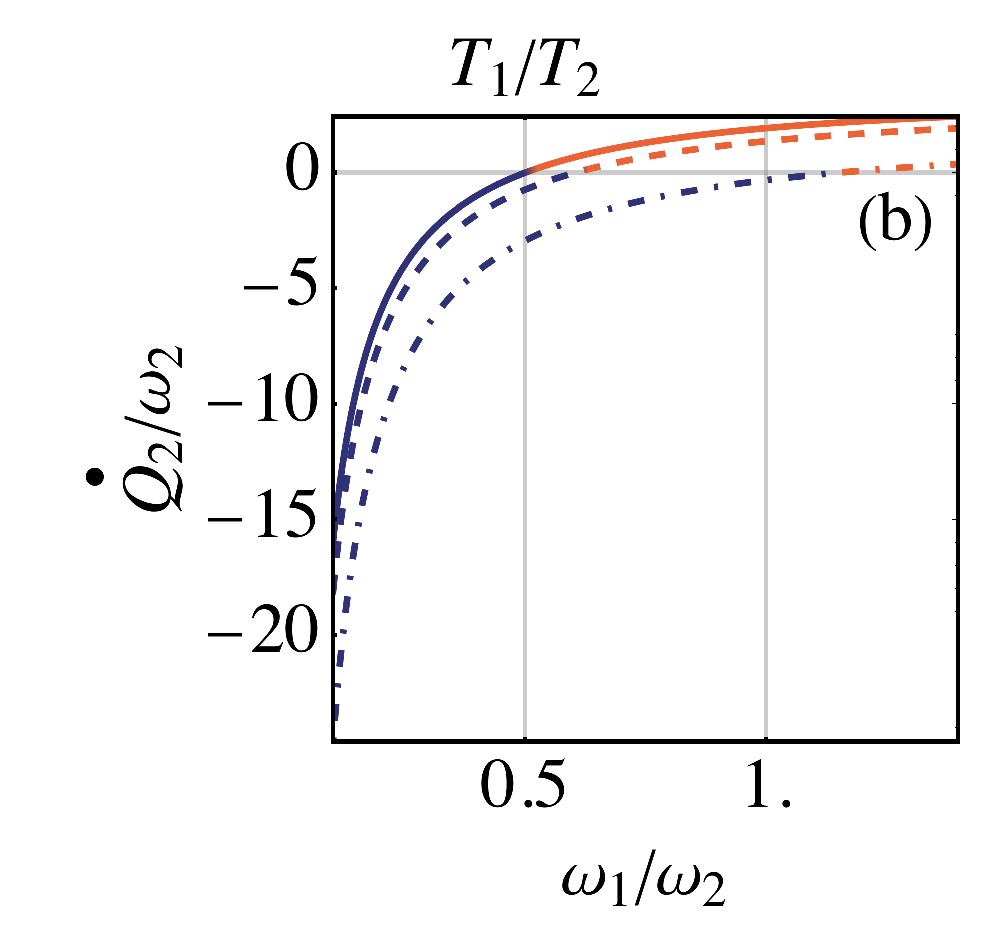}\;
\includegraphics[width=0.19\textwidth]{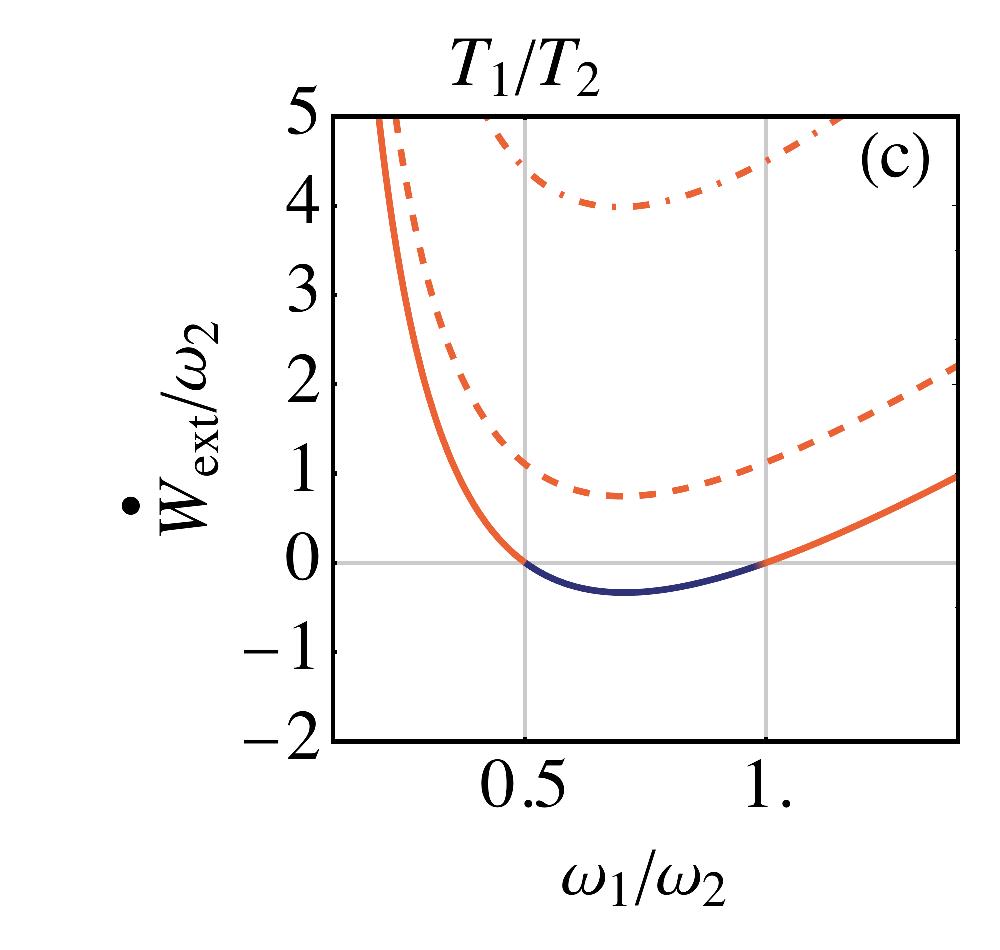}\;
\includegraphics[width=0.19\textwidth]{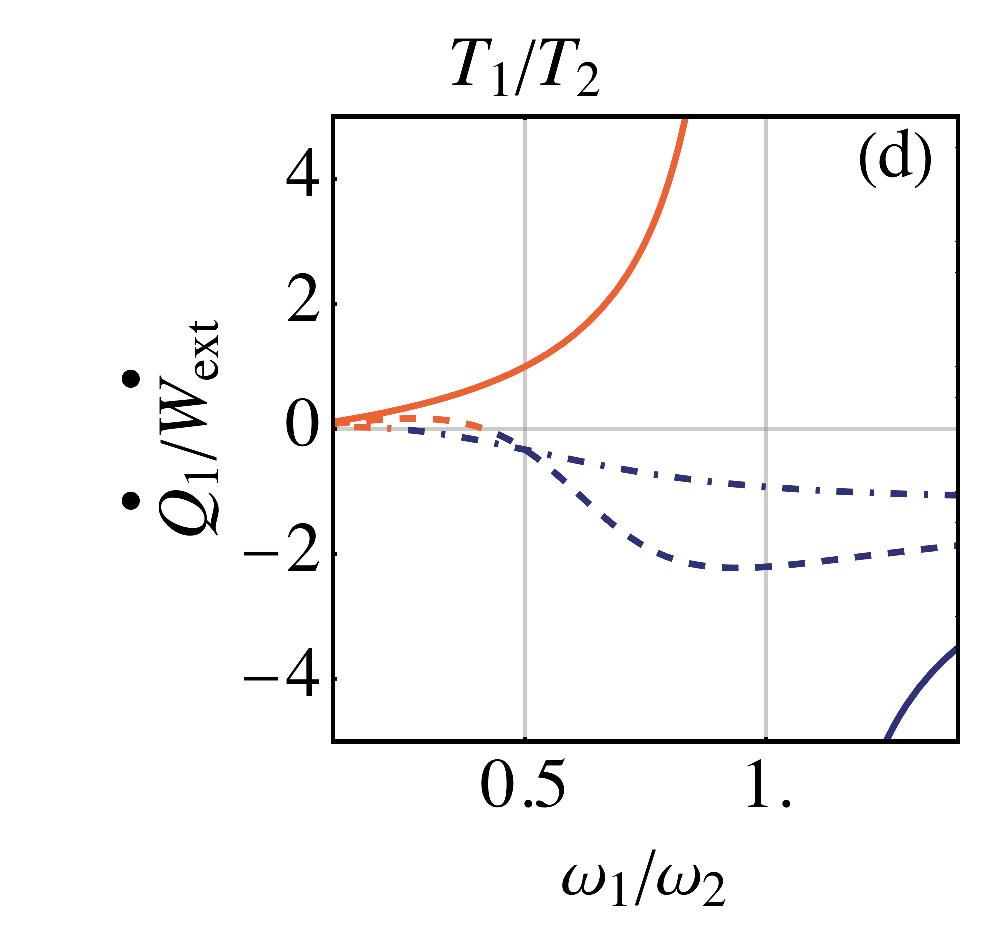}\;
\includegraphics[width=0.19\textwidth]{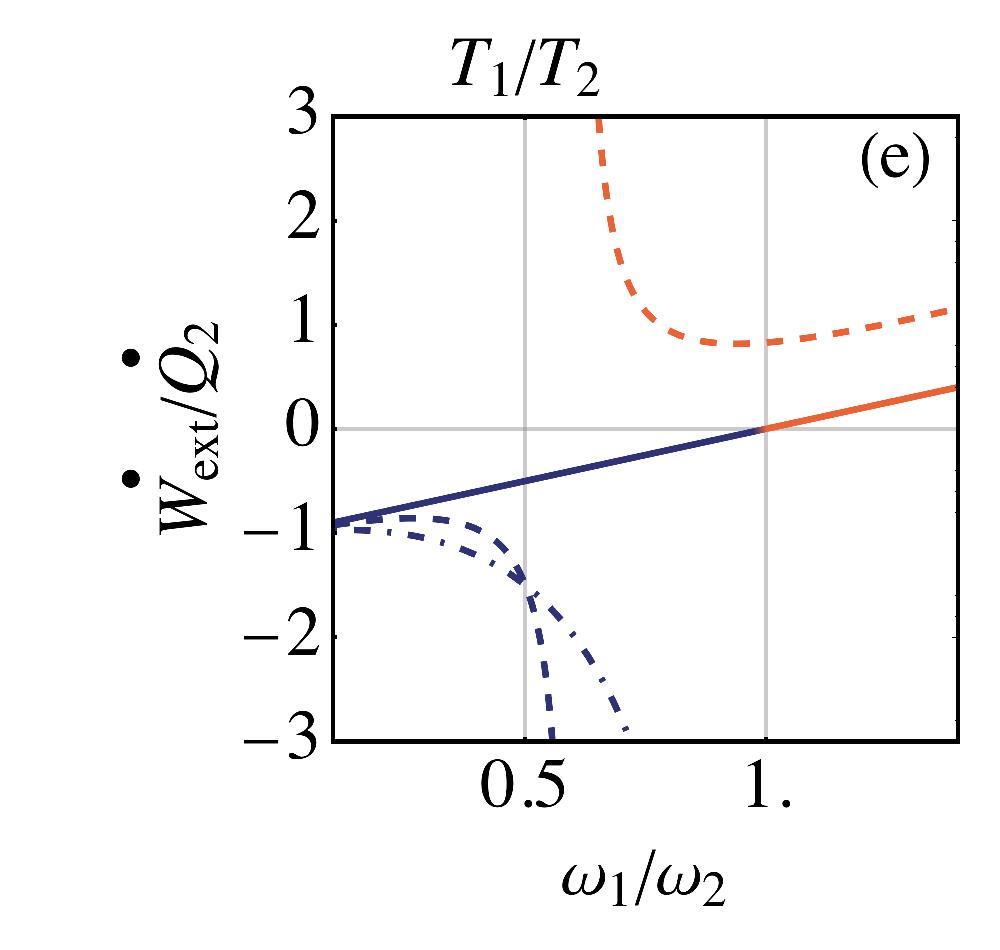}
\caption{\label{fig:2QHOs_again}
Similar to Fig.~\ref{fig:2QHOs}, but examining  the influence of the counter-rotating term $\eta$ in the Hamiltonian~(\ref{N_QHOs_HS}). 
The solid curves are the same as Fig.~\ref{fig:2QHOs}. 
The dashed curves correspond to Eqs.~(\ref{CR_Q1})-(\ref{CR_W}) for $\eta = 0.3 \omega_2$ (dashed) and $\eta = 0.6 \omega_2$ (dot-dashed). 
The curves were plotted for $\epsilon^2/\gamma = \omega_2$, assuming $T_1/T_2 = 1/2$.
The vertical lines correspond to $\omega_1/\omega_2 =T_1/T_2$ and $\omega_1=\omega_2$. 
}
\end{figure*}

%
%
\section{\label{sec:qho}Quantum Harmonic Oscillators}
%
%

We consider once again the problem studied in Sec.~\ref{sec:ss}, but we now make the following generalizations. 
First, we assume to have a one-dimensional chain of $N$, instead of 2, oscillators, with the first and last coupled to local baths. 
Second, we consider a more general type of interaction between them, so that we take the system Hamiltonian to be 
\begin{equation}\label{N_QHOs_HS}
H_S = \sum\limits_{i=1}^N \omega_i a_i^\dagger a_i + \sum\limits_{i=1}^{N-1} \bigg\{ \epsilon (a_i^\dagger a_{i+1} + a_{i+1}^\dagger a_i) + 
\eta (a_i^\dagger a_{i+1}^\dagger + a_{i+1} a_i) \bigg\}
\end{equation}
When $\eta = 0$ we recover the excitation-conserving tight-binding model. 
Conversely,  $\eta = \epsilon$ corresponds to  a position-position coupling $x_i x_{i+1}$. 
The total system evolves according to the master equation 
\begin{equation}
\frac{\ud \rho_S}{\ud t} = -i [H_S, \rho_S] + D_1(\rho_S) + D_N(\rho_S),
\end{equation}
where $D_i(\rho_S)$ is given in Eq~(\ref{qhos_Di}). 

Using Eq.~(\ref{heat_rate}) we find that the heat rate to the first and last environments will be 
\begin{equation}\label{2QHOs_gen_heat_rate}
\dot{Q}_i = \gamma \omega_i (n_i - \langle a_i^\dagger a_i \rangle), \qquad i = 1,N.
\end{equation}
The work rate, on the other hand, is found using Eq.~(\ref{work_rate}) and reads
\begin{IEEEeqnarray}{rCl}
\dot{W} &=& \frac{\gamma\epsilon}{2} \langle a_1^\dagger a_2 + a_1 a_2^\dagger + a_{N}^\dagger a_{N-1} + a_N a_{N-1}^\dagger \rangle 
\nonumber \\[0.2cm]
&&+ \frac{\gamma \eta}{2} \langle a_1 a_2 + a_1 a_2 + a_{N} a_{N-1} + a_N a_{N-1} \rangle.
\label{2QHOs_gen_work_rate}
\end{IEEEeqnarray}
If $N = 2$ and $\eta = 0$ this reduces to Eq.~(\ref{work_cost_2QHOs}).

\subsection{Degradation of the engine's operation due to counter-rotating terms }

As a first application, we return to the case of $N =2$ oscillators but generalize the results of Sec.~\ref{sec:ss} to include the counter-rotating terms $\eta$. 
We then show that, interpreting our steady-state as a heat engine, as in Fig.~\ref{fig:2QHOs}, the presence of this term only degrades the machine's operation. 
We once again use the Lyapunov equation technique of Appendix~\ref{sec:lyap}. 
Although the steady-state may be trivially found numerically, it turns out that the analytical expression for the covariance matrix becomes extremely cumbersome when $\eta \neq 0$. 
For the purpose of illustration, we therefore present here results that are valid when the bath coupling $\gamma$ is much larger than all other energy scales in the problem. 
In this case we find the following simple expressions for the heat rates~(\ref{2QHOs_gen_heat_rate}) and the work rate~(\ref{2QHOs_gen_work_rate}):
\begin{eqnarray}
\label{CR_Q1}
\dot Q_1 &=& \frac{2\omega_1}{\gamma} \bigg[ \epsilon^2 (n_1 - n_2) - \eta^2 (n_1 + n_2 + 1)\bigg],
\\
\label{CR_Q2}
\dot Q_2 &=&- \frac{2\omega_2}{\gamma} \bigg[ \epsilon^2 (n_1 - n_2) + \eta^2 (n_1 + n_2 + 1)\bigg],
\\
\label{CR_W}
\dot W_{\rm ext} &=& - \frac{2}{\gamma}\bigg[ \epsilon^2 (n_1 - n_2) (\omega_1 - \omega_2) + 
\\
&-&\eta^2 (\omega_1 + \omega_2)(n_1 + n_2 + 1)\bigg].
\nonumber
\end{eqnarray}
If $\eta = 0$ we recover Eqs.~(\ref{2qhos_heat_rate_1_explicit})-(\ref{2qhos_work_rate_explicit}), provided we also approximate $\Delta \simeq \gamma$. 

The influence of $\eta$ in these thermodynamic quantities is shown in Fig.~\ref{fig:2QHOs_again}, which compares the results with those of Fig.~\ref{fig:2QHOs} corresponding to $\eta = 0$.
As can be seen, the region in which the machine operates as a refrigerator or an engine is severely reduced by the presence of $\eta$. 
Moreover, as can be easily verified from Eqs.~(\ref{CR_Q1})-(\ref{CR_W}), in the case $\eta = \epsilon$ of a position-position coupling all of these operation regimes become inaccessible. 
This type of behavior has already been observed experimentally in the case of opto-mechanical systems, in which the interplay between $\eta$ and $\epsilon$ can be tuned by appropriately choosing the pump's sideband. 
And indeed, what is observed, is that since the $\eta$ interaction tends to entangle the two oscillators (since it leads to an evolution described by a two-mode squeezing operator), it has the general tendency of heating them up \cite{Wang2013,Woolley2014,Ockeloen-Korppi2018}, which is precisely what is observed in Fig.~\ref{fig:2QHOs_again}.

\subsection{Analysis for a chain of $N$ oscillators }

\begin{figure*}
\centering
\includegraphics[width=0.19\textwidth]{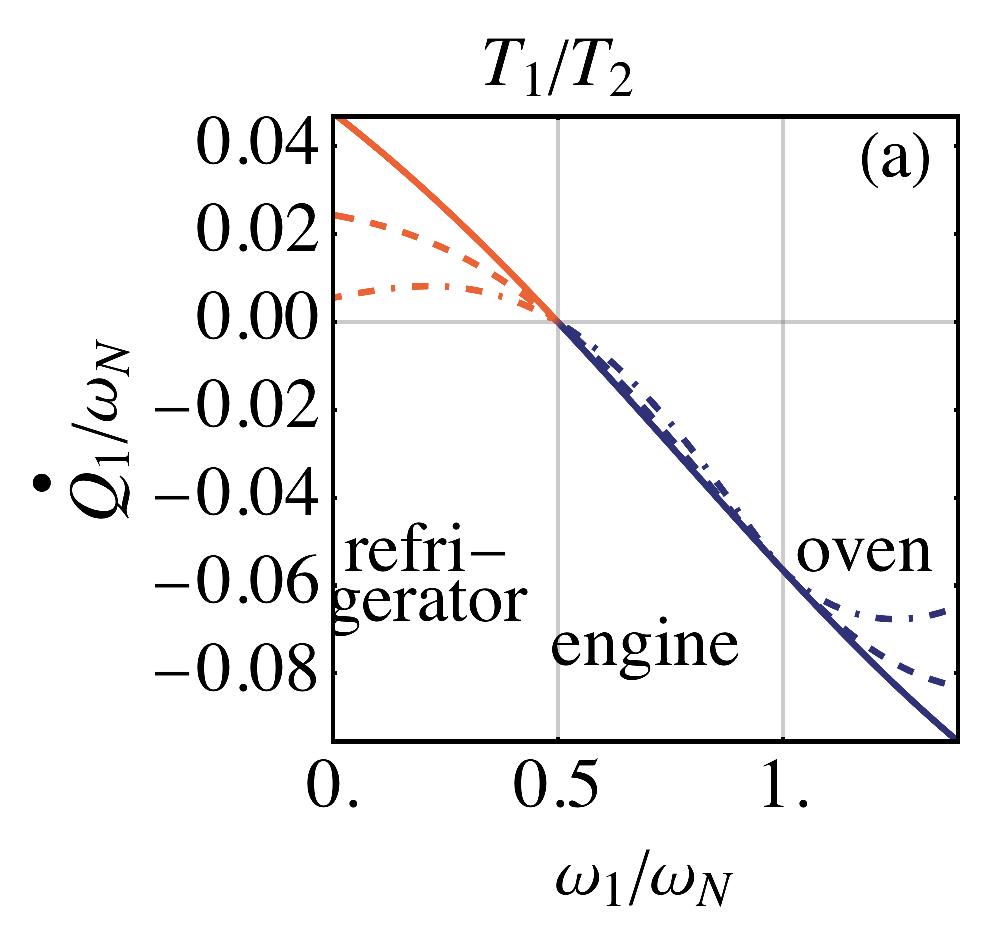}\;
\includegraphics[width=0.19\textwidth]{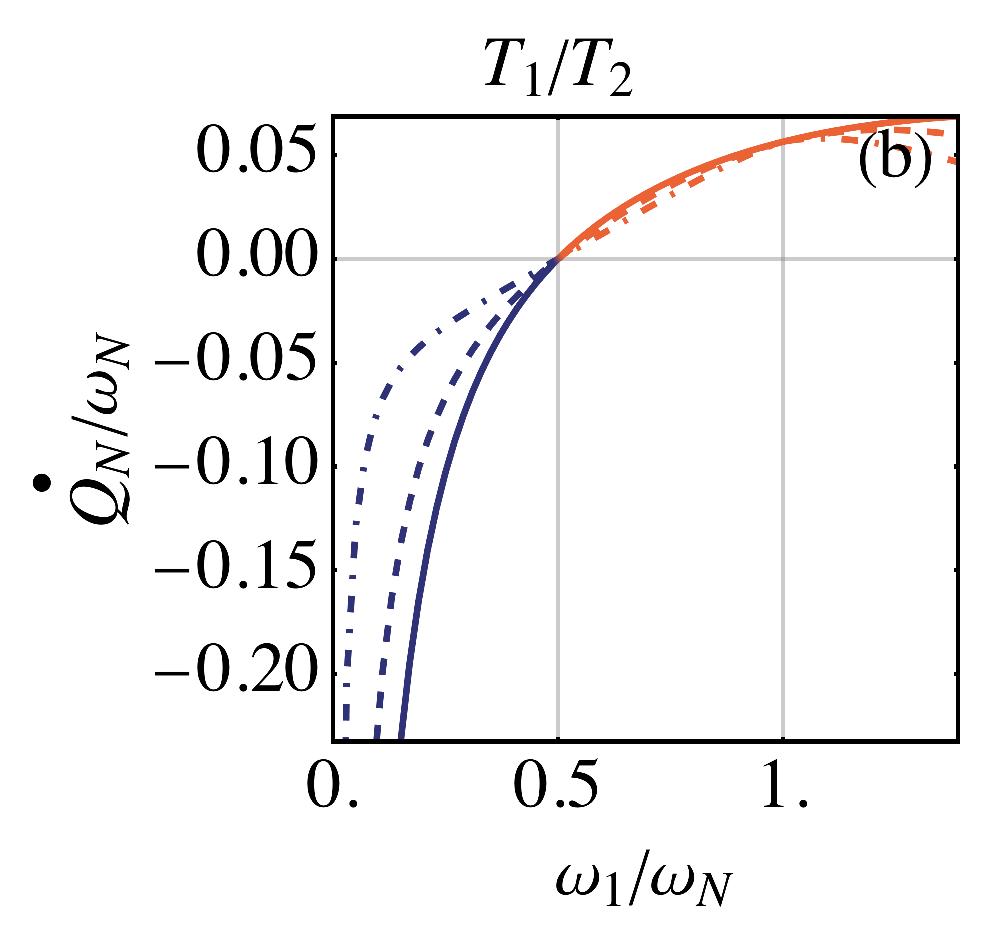}\;
\includegraphics[width=0.19\textwidth]{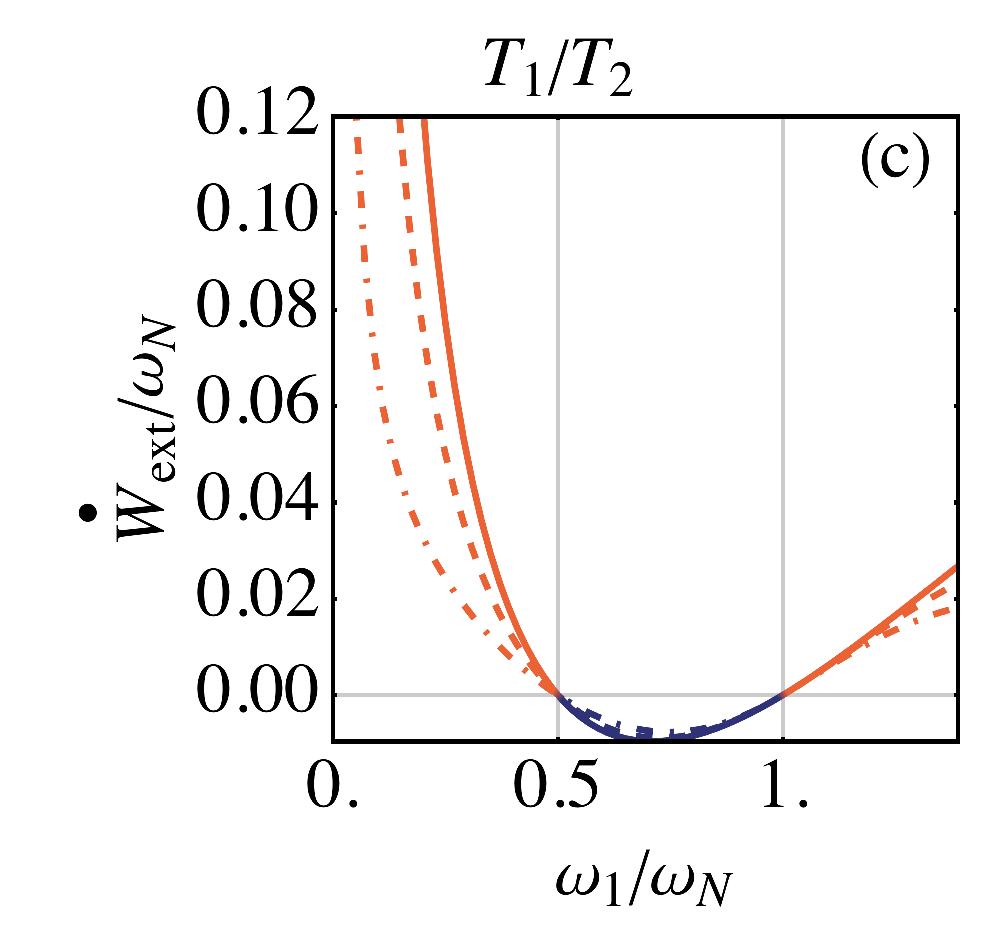}\;
\includegraphics[width=0.19\textwidth]{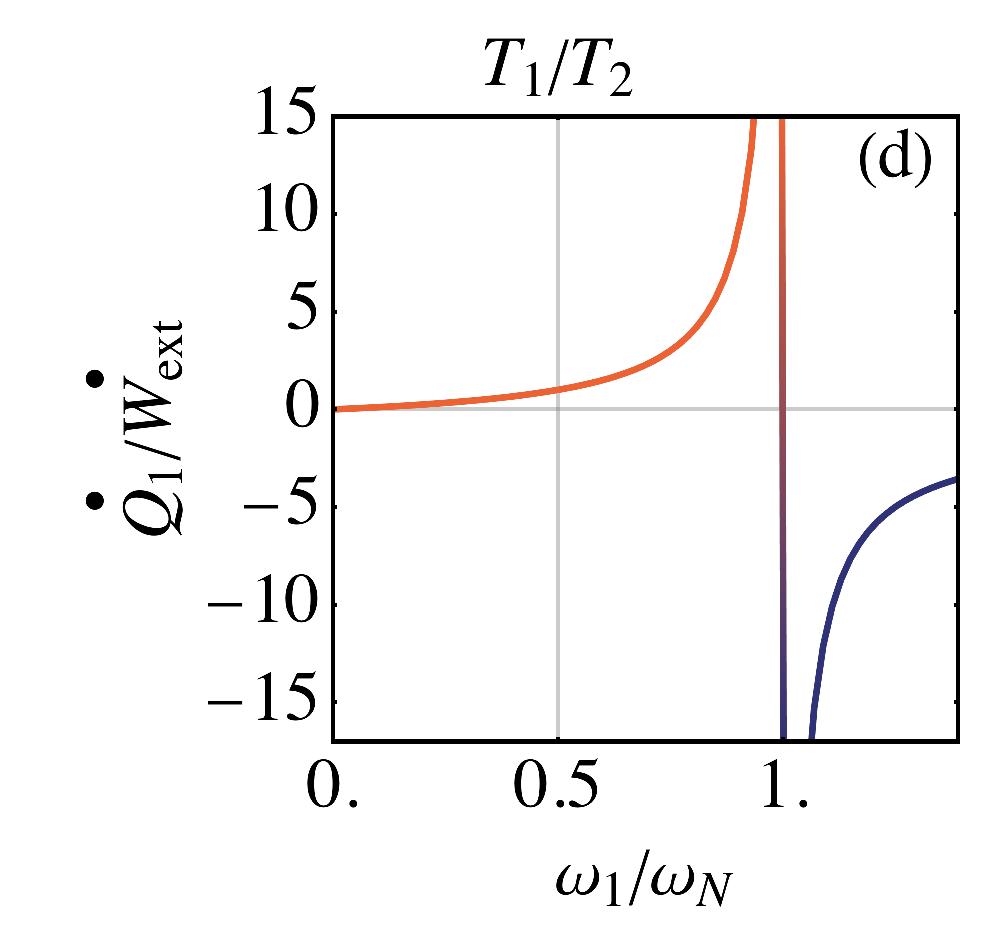}\;
\includegraphics[width=0.19\textwidth]{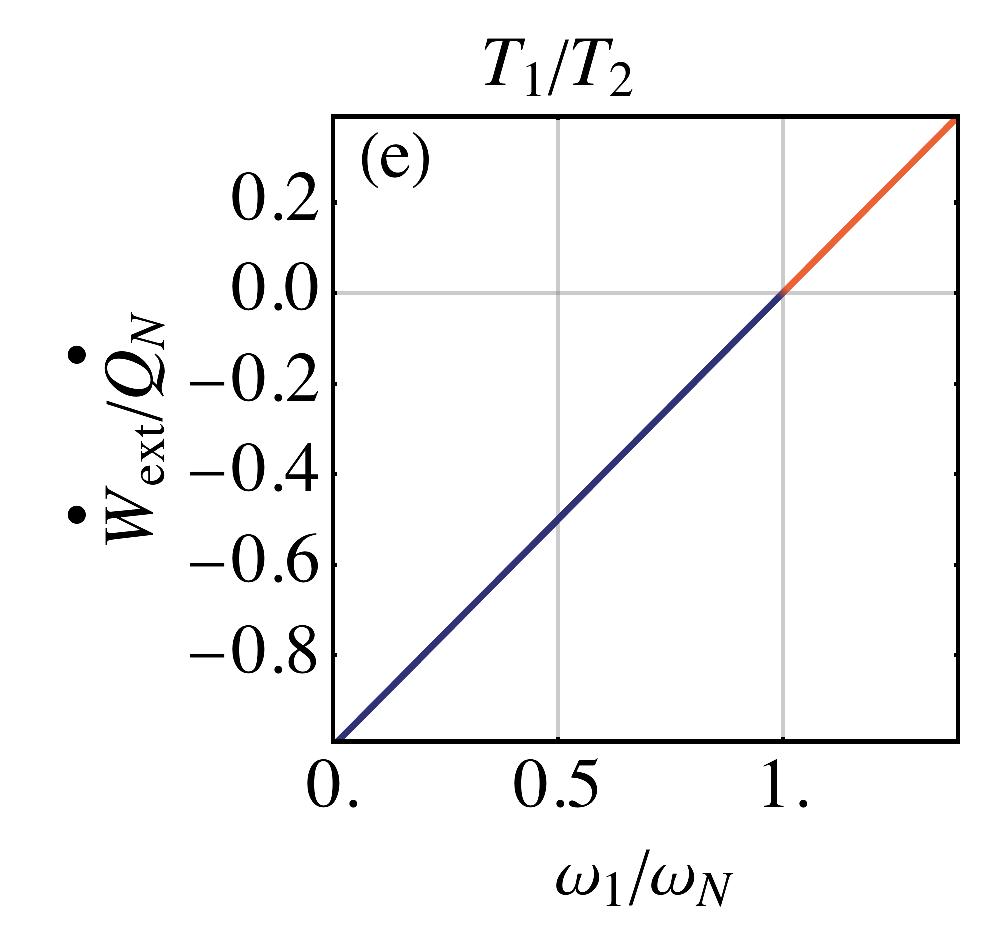}
\caption{\label{fig:NQHOs}
Similar to Fig.~\ref{fig:2QHOs}, but for a chain of $N  = 2$, 5 and 20 oscillators (solid, dashed and dot-dashed respectively) following a linear frequency profile [Eq.~(\ref{frequency_profile})].
The parameters were $\gamma = 2 \omega_N$, $\epsilon = 0.25 \omega_N$, $\eta = 0$ and $T_1/T_2 = 1/2$.
}
\end{figure*}

We next carry on a numerical analysis for the case of a chain of $N$ oscillators, assuming for simplicity that $\eta = 0$. 
The steady-state in this case is obtained by solving the Lyapunov equation numerically (Appendix~\ref{sec:lyap}). 
A set of results is presented in Fig.~\ref{fig:NQHOs} for the case of a linear function profile 
\begin{equation}\label{frequency_profile}
\omega_i = \frac{(N-i)\omega_1 + (i-1) \omega_N}{N-1}.
\end{equation}

As can be seen, in general the thermodynamic quantities depend on the system size.
However, the ratios between heat and work are independent of size and only depend on the frequencies $\omega_1$ and $\omega_N$ of the initial and final oscillators. 
This can be proven rigorously after writing the continuity equation for the number of excitations of each oscillator using techniques similar to those employed in Ref.~\cite{Asadian2013,Landi2014b}:
\begin{eqnarray}
\label{eq:cont1}
\frac{d\langle a^\dagger_1 a_1\rangle}{dt} &=&-i\epsilon\langle a^\dagger_1a_2-a_1a^\dagger_2\rangle+ \gamma \left(n_1-\langle a^\dagger_1 a_1\rangle\right),
\\
\frac{d\langle a^\dagger_i a_i\rangle}{dt} &=&-i\epsilon\langle a^\dagger_ia_{i+1}-a_ia^\dagger_{i+1}\rangle, \quad i=2,\dots,N-1
\\
\frac{d\langle a^\dagger_N a_N\rangle}{dt} &=&i\epsilon\langle a^\dagger_{N-1}a_N-a_{N-1}a^\dagger_N\rangle+ \gamma \left(n_N-\langle a^\dagger_N a_N\rangle\right).
\label{eq:cont2}
\end{eqnarray}

Hence, at steady state we have:
\begin{equation}
\gamma \left(n_1-\langle a^\dagger_1 a_1\rangle\right)=- \gamma \left(n_N-\langle a^\dagger_N a_N\rangle\right)
\end{equation}
and replacing this in the expressions for the heat flows~\eqref{2QHOs_gen_heat_rate} we obtain:
\begin{equation}
\dot Q_1 =\gamma\omega_1 \left(n_1-\langle a^\dagger_1 a_1\rangle\right) = -\frac{\omega_1}{\omega_N} \dot Q_N.
\end{equation}
Finally, using the first law of thermodynamics: $\dot W_{\rm ext}+\dot Q_1+\dot Q_N=0$ we obtain:
\begin{equation}
\label{eq:ratioschain}
\frac{\dot Q_1}{\dot W_{\rm ext}} = \frac{\omega_1}{\omega_N-\omega_1};
\quad
\frac{\dot W_{\rm ext}}{\dot Q_N} = -1+\frac{\omega_1}{\omega_N}.
\end{equation}
We remark that these ratios coincide with those of Otto thermal machines and refrigerators operating with a single harmonic oscillator driven between frequencies $\omega_1$ and $\omega_N$. Moreover, the universal ratios \eqref{eq:ratioschain} would hold if we replace the linear chain with a harmonic lattice of arbitrary geometry or replace the quantum harmonic oscillators with coupled qubits for which a continuity equation equivalent to Eqs.~\eqref{eq:cont1}-\eqref{eq:cont2}.

\section{Summary and Conclusions}
\label{sec:conclusions}
In this paper we have systematically analysed the external work required for local master equations based on the repeated interaction model. Specifically, we have provided explicit expressions, Eq.~\eqref{heat_rate} and~\eqref{work_rate}, for the relevant thermodynamics quantities as a function of the reservoir jump operators for a generic master equation in Lindblad form (with time independent rates).
We have found that, at steady state, this external work is directly related to the breaking of global detailed balance stemming from the internal system-system interactions.
Our analysis is general and can be applied to any array of interacting $d-$level systems. 

For the purpose of illustration, we have chosen to provide a detailed analysis of a chain of harmonic oscillators described by local master equations, a problem which has attracted considerable attention from the quantum thermodynamics and open quantum systems communities. 
We find that the heat transport through the chain is dramatically affected by the boundary driving and can even be inverted, leading to a flow of heat from the cold to the hot bath. 
However, this does not violate any thermodynamic law as this regime requires introducing positive work into the system, the latter effectively realising a quantum refrigerator.

Our findings therefore provide clear evidence that in order to be consistent with thermodynamics, one must take into consideration the microscopic model associated with a particular dynamical equation. 
Moreover, for a given process, complete positivity ensures that a quantum microscopic model can always be found. 
Hence, by taking into account all sources of energy exchange, we prove that the thermodynamic laws will always be satisfied.

\acknowledgements
We thank D. Alonso, F. Barra, M. Campisi, J. P. Paz for discussions.
GTL acknowledges the support from the S\~ao Paulo Research Foundation, under grant number 2016/08721-7. GDC and MA acknowledge the KITP program ``Thermodynamics of quantum systems: Measurement, engines, and control'' 2018, where part of this work has been done. This research was supported in part by the National Science Foundation under Grant No. NSF PHY-1748958. AJR acknowledges support from CONICET, ANPCyT and UBACyT. AF acknowledges support from the EPSRC project EP/P00282X/1.

\appendix

\section{\label{sec:lyap}The Lyapunov equation}

We discuss here how to compute the steady-state of Gaussian preserving master equations. 
For concreteness, we focus on the problem discussed in Sec.~\ref{sec:ss}, more specifically Eq.~(\ref{eq:MElocaloscillators}). 
The extension to multiple oscillators (Sec.~\ref{sec:qho}) is straightforward. 
Since the master equation is Gaussian preserving, the steady-state must necessarily be Gaussian.
Hence, it is completely determined by the  second moments of the oscillators' positions and momenta, arranged for convenience in the vector:
\begin{equation}
Y^T=\left(x_1,p_1,x_2,p_2\right).
\end{equation}
We introduce the drift matrix containing the couplings and decay constants:
\begin{equation}
K=
\left(
\begin{array}{cccc}-\gamma/2 & 1 & 0 & \lambda \\
-\omega_1^2 & -\gamma/2 & -\mu & 0 \\
0 & \lambda & -\gamma/2 & 1 \\
-\mu & 0 & -\omega_2^2 & -\gamma/2
\end{array}
\right)
\end{equation}
where:
\begin{equation}
\lambda=\frac{2\epsilon}{\sqrt{\omega_1\omega_2}},\quad\quad
\mu=2\epsilon \sqrt{\omega_1\omega_2}
\end{equation}
and the system's covariance matrix $V$ with entries:
\begin{equation}
V(i,j)=\frac 12 \langle\;\{ Y(i),Y(j)\} \;\rangle
\end{equation}
 the average being done on the state of the system. By calculating the equation of motion for all the elements of the covariance matrix, we obtain the following Lyapunov equation for the covariance matrix:
\begin{equation}
\label{eq:2oscme}
\dot V=KV+VK^T+D
\end{equation}
where 
\begin{eqnarray*}
D=&&\diag\left[\frac{\gamma(2n_1+1)}{2 \omega_1},\frac{\gamma  \omega_1(2n_1+1)}{2},  \right .
\\
&& \left . \frac{\gamma(2n_2+1)}{2 \omega_2},\frac{\gamma  \omega_2(2n_2+1)}{2}\right].
\end{eqnarray*}

The steady-state is then determined by the algebraic equation
\begin{equation}
KV+VK^T+D=0.
\end{equation}
In absence of coupling between the first and second oscillator, $\epsilon=0$, the stationary state of each oscillator is thermal at its respective temperature:
\begin{equation}
V=
\left(
\begin{array}{cccc}
\frac{2n_1+1}{2\omega_1} & 0 & 0 & 0 \\
0 &\frac{(2n_1+1) \omega_1}{2} & 0 & 0 \\
0 & 0 & \frac{2n_2+1}{2 \omega_2} & 0 \\
0 & 0 & 0 & \frac{(2n_2+1) \omega_2}{2}
\end{array}
\right)
\end{equation}

When $\epsilon \neq 0$, one finds instead the following expression for the entries of $V$ (recall that $V_{ij}=V_{ji}$):
\begin{eqnarray}
V_{11}&=&\frac{1}{2
    \omega_1}\left(\frac{4 \epsilon
   ^2 (n_2-n_1)}{\Delta^2}+2 n_1+1\right),
   \\
  V_{22}&=& \frac{1}{2} 
   \omega_1 \left(\frac{4 \epsilon ^2
   (n_2-n_1)}{\Delta^2}+2
   n_1+1\right),
   \\
  V_{33}&=& \frac{1}{2
    \omega_2}\left(\frac{4 \epsilon
   ^2 (n_1-n_2)}{\Delta^2}+2 n_2+1\right),   \\
 V_{44}&=& \frac{1}{2} 
   \omega_2 \left(\frac{4 \epsilon ^2
   (n_1-n_2)}{\Delta^2}+2
   n_2+1\right),
   \\
  V_{13}&=& \frac{ \epsilon  (n_1-n_2)
   (\omega_1-\omega_2)}{
   \sqrt{\omega_1 \omega_2}
   \Delta^2},
   \\
  V_{14}&=& \frac{ \gamma
   \omega_2 \epsilon 
   (n_2-n_1)}{\sqrt{\omega_1
   \omega_2} \Delta^2},
   \\
  V_{23}&=& \frac{ \gamma \omega_1 \epsilon  (n_1-n_2)}{\sqrt{\omega_1 \omega_2} \Delta^2},
   \\
  V_{24}&=& \frac{  \epsilon 
   (n_1-n_2) (\omega_1-\omega_2) \sqrt{\omega_1 \omega_2}}{\Delta^2},
   \\
    V_{12}&=&V_{34} =0.
\end{eqnarray}
which are tantamount to Eqs.~(\ref{2QHOs_occ1})-(\ref{2QHOs_occ2}).

%
%
\section{\label{sec:details}Details of the repeated interactions derivation}
%
%

In this section we provide additional details on how to establish the thermodynamic quantities within the method of repeated interactions.  
A single interaction stroke between system and environment is described by the unitary map
\[
\rho_{SE}' = e^{-i \tau H_\text{tot}} \; \rho_S \rho_E \; e^{i \tau H_\text{tot}},
\]
where $H_\text{tot}$ is given in Eq.~(\ref{Htot}) (with $V_i \to V_i/\sqrt{\tau}$, as discussed in the main text). 
Hence, the evolution of any observable $\mathcal{O}$ (from either the system or from the environments) due to this interaction is given by 
\begin{equation}\label{expectation_value_evolution}
\langle \mathcal{O} \rangle' = \langle e^{i \tau H_\text{tot}}\; \mathcal{O} e^{-i \tau H_\text{tot}}  \rangle,
\end{equation}
where the unprimed average in the right-hand side is computed with respect to the initial state before the evolution. 
Using this idea we find that the change in the energy of the system $H_S$ and of each environment $H_{E_i}$ is up to order $\tau$
\begin{IEEEeqnarray}{rCl}
\Delta H_S &=& - \frac{\tau}{2} \sum\limits_{i=1}^N \langle [V_i, [V_i, H_S]] \rangle, 
\\[0.2cm]
\Delta H_{E_i} &=& - \frac{\tau}{2}  \langle [V_i, [V_i, H_{E_i}]] \rangle.
\end{IEEEeqnarray}
Carrying out the computations, using Eq.~(\ref{Vi}), we find 
\begin{IEEEeqnarray}{rCl}
\Delta H_S &=& \frac{\tau}{2} \sum\limits_{i,k} g_{i,k}^2 \bigg\{ \langle A_{i,k}^\dagger A_{i,k} \rangle \langle L_{i,k} [H_S, L_{i,k}^\dagger] + [L_{i,k}, H_S] L_{i,k}^\dagger \rangle \nonumber
\\[0.2cm]
&&\qquad \quad + \langle A_{i,k} A_{i,k}^\dagger \rangle \langle L_{i,k}^\dagger [H_S, L_{i,k}] + [L_{i,k}^\dagger, H_S] L_{i,k} \rangle \bigg\}
\nonumber\\[0.2cm]
\Delta H_{E_i} &=&- \tau \sum\limits_k g_{i,k}^2 \omega_{i,k} \bigg\{ \langle A_{i,k}^\dagger A_{i,k} \rangle \langle L_{i,k} L_{i,k}^\dagger \rangle - \langle A_{i,k} A_{i,k}^\dagger \rangle \langle L_{i,k}^\dagger L_{i,k} \rangle \bigg\}.
\IEEEeqnarraynumspace\nonumber
\end{IEEEeqnarray}
Here the averages over the environment operators are always computed with respect to the same state $\rho_E$, whereas the average over system operators are computed over the instantaneous state $\rho_S(t)$. 

Thus, identifying the transition rates $\gamma_{i,k}^\pm$ in Eq.~(\ref{gamma_pm}), we may then divide both sides by $\tau$ and take the limit $\tau\to 0$, which yields
\begin{IEEEeqnarray}{rCl}
\frac{\ud \langle H_S \rangle}{\ud t} &=&\frac{1}{2} \sum\limits_{i,k} \bigg\{ \gamma_{i,k}^+ \langle L_{i,k} [H_S, L_{i,k}^\dagger] + [L_{i,k}, H_S] L_{i,k}^\dagger\rangle +
\\[0.2cm]
&&\qquad \quad+ \gamma_{i,k}^- \langle L_{i,k}^\dagger [H_S, L_{i,k}] + [L_{i,k}^\dagger, H_S] L_{i,k} \rangle \bigg\}.
\nonumber
\end{IEEEeqnarray}
Moreover, with respect to the change in energy of the environments, we define the heat rates as $\dot Q_i = - \Delta H_{E_i}/\tau$, which then yields precisely Eq.~(\ref{heat_rate}). 

The heat rate may now be found in two ways. 
First, using Eq.~(\ref{HS}) and noting that 
\begin{equation}
[H_S, L_{i,k}] = -\omega_{i,k} L_{i,k} + [H_I, L_{i,k}], 
\end{equation}
we see that $\ud \langle H_S \rangle/\ud t$ can be split into two terms, one of which is precisely  $\sum_i \dot Q_i$. 
Hence, the remainder must be attributed to the work that has to be performed, which yields precisely Eq.~(\ref{work_rate}). 

Instead, another way of defining the work rate is by noting that since one must decouple the system from the environments at each stroke, the total Hamiltonian is actually time dependent. 
That is, if we focus on just a single interaction stroke, then instead of Eq.~(\ref{Htot}), the Hamiltonian for this interaction is more appropriately defined as 
\begin{equation}
H_\text{tot} = H_S + H_E + \frac{\lambda(t)}{\sqrt{\tau}} \sum\limits_{i=1}^N V_i,
\end{equation}
where $\lambda(t)$ has the value 1 when $t \in [n \tau, (n+1)\tau]$ and zero otherwise. 
Since the global S+E interaction is unitary, work can be unambiguously defined as 
in Eq.~(\ref{work_single_stroke}). 
Carrying out the integration one then finds
\begin{equation}
\delta W = \sum\limits_{i=1}^N \frac{\langle V_i \rangle - \langle V_i \rangle'}{\sqrt{\tau}},
\end{equation}
where, once again, the primed expectation value refers to the state after the interaction. 
Using once again Eq.~(\ref{expectation_value_evolution}) for evaluating this expectation value, we then find 
\begin{equation}\label{delta_W_sum_terms}
\delta W = \frac{\tau}{2} \sum\limits_{i=1}^N \langle [V_i, [V_i, H_S + H_{E_i}]]\rangle = \Delta H_S + \sum\limits_{i=1}^N \Delta H_{E_i}
\end{equation}
Hence, taking again the limit $\tau \to 0$, one identifies the work rate as $\dot W = \frac{\ud \langle H_S \rangle}{\ud t} - \sum_i \dot Q_i$. 

%
%
\section{\label{sec:prod}Entropy production in local master equations}
%
%

In this appendix we discuss how to express the second law of thermodynamics within the repeated interactions scheme. 
Since the interaction strokes only last for a small time $\tau$, one may neglect any potential bath-bath correlations that may appear when a system is interacting with multiple environments. 
Hence, following~\cite{Strasberg2016}, we  can define the entropy production in a single stroke as 
\begin{equation}\label{Sigma_info}
\Sigma := \mathcal{I}(S:E),
+ S(\rho_E(\tau) || \rho_E^\text{th}) \geq 0.
\end{equation}
where
\begin{equation}
\mathcal{I}(S:E) = \Delta S_S + \Delta S_E \simeq \Delta S_S + \sum\limits_{i=1}^N \Delta S_{E,i},
\end{equation}
 is the mutual information developed between the system and all the environments, 
 with $\Delta S_S$ being the change in the entropy of the system and $\Delta S_{E_i}$  the change in the von Neumann entropy of environment $E_i$.
The second term in Eq.~(\ref{Sigma_info}), on the other hand,
\begin{equation}
S(\rho_E(\tau) ||\rho_E^\text{th}) \simeq \sum\limits_{i=1}^N S(\rho_{E,i}(\tau) ||\rho_{E,i}^\text{th}),
\end{equation}
is the quantum relative entropy between the final and initial states of each environment  [here $S(\rho||\sigma) = \tr(\rho \ln \rho - \rho \ln \sigma)$].
The positivity of $\Sigma$ then follows immediately from the positivity of  the mutual information and the quantum relative entropy.

Using Eq.~(\ref{delta_W_sum_terms}) one may then readily show that Eq.~(\ref{Sigma_info}) may be written as 
\begin{equation}
\Sigma = \Delta S_S - \sum\limits_{i=1}^N \beta_i \delta Q_i,
\end{equation}
Dividing by $\tau$ and taking the limit $\tau \to 0$ we may then obtain the usual expression for the entropy production rate
\begin{equation}\label{Pi_thermo}
\Pi = \frac{\ud S_S}{\ud t} - \sum\limits_{i=1}^N \beta_i \frac{\ud Q_i}{\ud t},
\end{equation}
with the last term representing the entropy flux rates to each environment. 

In the particular case where there is a single environment, or when all environments have the same temperature, we may use the first law to write this as 
\begin{equation}\label{Pi_thermo_2}
\Pi = \beta \left( \frac{\ud W}{\ud t} - \frac{\ud F}{\ud t}\right),
\end{equation}
where $F_S = \langle H_S \rangle - T S_S$ is the free energy of the system. 
Thus, in this case we recover another well known expression for the entropy production rate.

\bibliographystyle{apsrev4-1}
\bibliography{biblio}
\end{document}